\documentclass[aps,prd,amsmath,amssymb,twocolumn,preprintnumbers,nofootinbib]{revtex4-2}
\usepackage[T1]{fontenc}
\usepackage{times}
\DeclareMathAlphabet{\MATHIT}{OT1}{ptm}{m}{it}
\DeclareSymbolFont{Letters}{OML}{ztmcm}{m}{it}
\DeclareSymbolFontAlphabet{\mathNormal}{Letters}
\usepackage[utf8]{inputenc}
\usepackage[english]{babel}
\usepackage{amsmath}
\usepackage{amsfonts}
\usepackage{amssymb}
\usepackage{amsthm}
\usepackage{enumitem}
\usepackage{dcolumn}
\usepackage{bm}
\usepackage{bbm}
\usepackage{cancel}
\usepackage{tikz}
\usetikzlibrary{calc,shapes,arrows,patterns,decorations.pathmorphing,positioning,automata}
\usepackage{tabularx}
\usepackage{booktabs}
\usepackage{tensor}
\usepackage{scalerel}
\usepackage{xcolor}
\usepackage{mathtools}
\usepackage{float}
\usepackage{xfrac}
\usepackage[pdfborder={0 0 0}]{hyperref}
\definecolor{darkblue}{rgb}{0,0,.5}
\definecolor{darkgreen}{rgb}{0,0.5,.5}
\definecolor{darkyellow}{rgb}{0.5,0.5,0}
\definecolor{fhl}{rgb}{1,0,0}
\hypersetup{colorlinks=true, breaklinks=true, linkcolor=darkblue, menucolor=darkblue, urlcolor=darkblue, linktocpage=true}
\usepackage{geometry}
\usepackage{textpos}
\usepackage{relsize}
\usepackage{balance}
\usepackage[titletoc,title]{appendix}
\allowdisplaybreaks
\usepackage{slashed}
\usepackage{natbib}
\usepackage[export]{adjustbox}

\usepackage{siunitx}  

\newcommand{\be}{\begin{equation}}
\newcommand{\ee}{\end{equation}}
\newcommand{\beq}{\begin{eqnarray}}
\newcommand{\eeq}{\end{eqnarray}}

\newcommand{\bmp}{\noindent\begin{minipage}{16cm}}
\newcommand{\emp}{\end{minipage}\vskip 7mm} 

\def\lsim{\mathrel{\rlap{\lower4pt\hbox{\hskip1pt$\sim$}}
    \raise1pt\hbox{$<$}}}                
\def\gsim{\mathrel{\rlap{\lower4pt\hbox{\hskip1pt$\sim$}}
    \raise1pt\hbox{$>$}}}                

\let\originalleft\left
\let\originalright\right
\renewcommand{\left}{\mathopen{}\mathclose\bgroup\originalleft}
\renewcommand{\right}{\aftergroup\egroup\originalright}
\newcommand{\SU}[1]{\operatorname{SU}\left(#1\right)}
\newcommand{\of}[1]{\left(#1\right)}

\newcommand{\ssof}[1]{(#1)}

\newcommand{\cof}[1]{\left\{#1\right\}}

\newcommand{\avof}[1]{\left\langle #1\right\rangle}

\renewcommand*\[{\begin{equation}}
\renewcommand*\]{\end{equation}}

\definecolor{newgreen}{RGB}{10,100,20}

\begin{document}


\title{Spectrum of SU(2) gauge theory at large number of flavors}

\author{Jarno Rantaharju}
\email{jarno.rantaharju@helsinki.fi}
\affiliation{Department of Physics \& Helsinki Institute of Physics,
P.O. Box 64, FI-00014 University of Helsinki}

\author{Tobias Rindlisbacher}
\email{tobias.rindlisbacher@helsinki.fi}
\affiliation{Department of Physics \& Helsinki Institute of Physics,
P.O. Box 64, FI-00014 University of Helsinki}
\author{Kari Rummukainen}
\email{kari.rummukainen@helsinki.fi}
\affiliation{Department of Physics \& Helsinki Institute of Physics,
P.O. Box 64, FI-00014 University of Helsinki}
\author{Ahmed Salami}
\email{ahmed.salami@helsinki.fi}
\affiliation{Department of Physics \& Helsinki Institute of Physics,
P.O. Box 64, FI-00014 University of Helsinki}
\author{Kimmo Tuominen}\email{kimmo.i.tuominen@helsinki.fi}
\affiliation{Department of Physics \& Helsinki Institute of Physics,
P.O. Box 64, FI-00014 University of Helsinki}

\begin{abstract}
We present a numerical study of the spectrum of an asymptotically non-free $\SU{2}$ gauge
theory with $N_f=24$ massive fermion flavors. For such large number of flavors, asymptotic
freedom is lost and the massless theory is governed by a gaussian fixed point at long distances.  If fermions are massive they decouple at low energy scales and the theory is confining.  We present a scaling law for the masses of the hadrons, glueballs and string tension as functions of fermion mass.  The hadrons become effectively heavy quark systems, with masses approximately twice the fermion mass, whereas the energy scale of the confinement, probed by e.g. the string tension, is much smaller and vanishes asymptotically as $m_\text{fermion}^{2.18}$.
Our results from lattice simulations are compatible with this behaviour.
\end{abstract}
\maketitle

\section{Introduction}\label{sec:intro}

Asymptotically free nonabelian gauge-fermion theories are a cornerstone of our
theoretical understanding of the elementary particle interactions of ordinary matter.
The first principle methodologies of lattice field theory to solve these theories
where preturbative methods are inapplicable are well established. Recently much effort has
been devoted to studies of gauge-fermion theories whose matter content facilitates the
existence of an infrared fixed point~\cite{Sannino:2004qp} as such theories are of
interest for beyond standard model phenomenology~\cite{Hill:2002ap,Dietrich:2005jn,Arbey:2015exa}.
On the lattice the properties of this type of theories have been studied for SU(2) gauge theory
with matter fields in the
fundamental~\cite{Karavirta:2011zg,Leino:2017lpc,Leino:2017hgm,Leino:2018qvq,Amato:2018nvj}
or adjoint~\cite{Hietanen:2008mr,Hietanen:2009az,DelDebbio:2008zf,DelDebbio:2009fd,DelDebbio:2010hu,Bursa:2011ru,DeGrand:2011qd,Rantaharju:2015yva,DelDebbio:2015byq} representation.
For related studies in SU(3) case, see e.g.~\cite{Fodor:2011tu,Hasenfratz:2014rna,Fodor:2015baa,Fodor:2017gtj,LatticeStrongDynamics:2018hun,Hasenfratz:2019dpr,Hasenfratz:2020ess} for fundamental representation
fermions and~\cite{Shamir:2008pb,DeGrand:2008kx,Fodor:2009ar,DeGrand:2010na,Fodor:2015zna} for two-index symmetric representation. Analyses have been extended also to SU(4) gauge group with
fermions in fundamental and higher representations~\cite{DeGrand:2012qa,DeGrand:2015lna}.

However, less is known of the precise dynamics of gauge theories with many
fermion flavors so that the theory is no longer asymptotically free. While such a
particle content is not directly relevant for the Standard Model of elementary particle
interactions, it poses an interesting challenge for general understanding of
gauge theory dynamics on one hand and on the development of computational methods on the other.

Broadly, the non-asymptotically free theories can be categorized as trivial or asymptotically
safe. The former means that the theory develops an ultraviolet (UV) cutoff and can be
interpreted consistently as a fundamental theory only as a free theory at short distances,
while the latter means that a non-perturbative
UV fixed point emerges and controls the short distance behavior of the theory.

In an earlier study~\cite{Leino:2019qwk}, we analysed this question in SU(2) gauge theory with 24
and 48 massless Dirac fermions by measuring the evolution of the coupling constant.
Our results suggested
that SU(2) gauge theory at these large numbers of fermions does not have an UV fixed point
but is in the category of trivial theories. The results on the running coupling were seen
to match well with the perturbative running in the infared (IR).

In this paper we continue the lattice investigation of SU(2) gauge theory with 24 flavors.
The aim is to complement our earlier results on the measurement of the coupling
by a computation of the spectrum of the theory, i.e. the determination of long distance
or infrared (IR) behavior, as the quark\footnote{For concreteness, we call the fermions ``quarks''.} mass $m_q$ is varied.  When the quark mass is non-vanishing the infrared behaviour of the theory changes: at energy scales smaller than the quark mass $m_q$, the quarks decouple and the physics is effectively that of confining SU(2) gauge theory with heavy quarks.  The spectrum of the theory includes two quark baryons, quark-antiquark mesons and glueballs, and the string tension is non-vanishing.  In terms of the renormalization group (RG) flow, the point $(g^2 = 0,m_q = 0)$ is an infrared fixed point, and $g^2$ and $m_q$ are irrelevant and relevant paramters, respectively.

We introduce scaling laws for the hadron masses and quantities probing confinement (string tension, glueballs) as the quark mass is reduced.  Our main result is that the hadrons behave as heavy quark systems, with masses close to $2m_q$, whereas the energy scale characterizing confinement is much smaller, and asymptotically vanishes proportional to $m_q^{2.18}$.  Our lattice measurements agree with these predictions, although the confinement scale turns out to be so small that we can only give upper limits to it.
%
%
Together with our earlier work, the present paper establishes a consistent picture
for the nonperturbative behaviors in this theory from IR to UV scales.

The paper is organized as follows: in Section \ref{sec:spectrum} we introduce the expected scaling of the relevant mass scales of the theory. In Section \ref{sec:latform} we
briefly describe the lattice formulation and in Section \ref{sec:results} we
describe the lattice measurements and the results. In Section \ref{sec:conclusion}
we conclude and ouline possibilities for further work.

\section{Expected mass spectrum}\label{sec:spectrum}

The solution of the 1-loop $\beta$-function for SU(2) gauge theory with $N_f$ massless quarks is
\be
  g^2(\mu) = \frac{1}{2\beta_0^{(N_f)} \log(\mu/\Lambda) }\,.
  \label{g1loop}
\ee
Let us consider the case $N_f=24$.  Now
$\Lambda = \Lambda_{\text{UV}}$ is the UV Landau pole, $\mu < \Lambda_{\text{UV}}$ and $\beta_0^{(24)} \approx -0.0549$.  Thus, when the energy scale $\mu\rightarrow 0$
the coupling constant $g^2$ vanishes, and the system is free in the infrared: there are no bound states, and the string tension vanishes.

However, if the quarks are massive the situation changes dramatically: when $\mu \ll m_q$, the quarks decouple and the system effectively becomes confining pure gauge SU(2) theory.  String tension is non-vanishing, and the mass spectrum includes glueballs, quark-antiquark mesons and two-quark baryons.

The energy scale of the confinement of SU(2) gauge theory sets the mass scale of glueballs and string tension.  We can make a rough estimate of the confinement scale with the 1-loop
running of the coupling in $N_f=24$ and $N_f=0$ theories, and setting the couplings equal at $\mu=m_q$:
\be
  g^2(\mu=m_q,N_f=24) = g^2(\mu=m_q,N_f=0)\,.
\ee
Let us call the $\Lambda$-parameter of the $N_f=0$ theory $\Lambda_\text{IR}$.  It is analogous to ``$\Lambda_{\text{QCD}}$'' of the pure gauge theory and is a proxy for the confinement energy scale. Alternatively, we could use a fixed large value of $g^2(\mu)$ to set the scale, and conclusions would remain unchanged.

Solving for $\Lambda_\text{IR}$ in terms of the quark mass and the UV scale $\Lambda_{\text{UV}}$ we obtain
\be
  \frac{\Lambda_{\text{IR}}} {\Lambda_{\text{UV}}} = \left(\frac{m_q}{\Lambda_{\text{UV}}}\right)
                ^{1 - \beta_0^{(24)}/\beta_0^{(0)}}
  \approx \left(\frac{m_q}{\Lambda_\text{UV}}\right)
                ^{2.18}\,.
\label{approxLambda0}
\ee
Thus, the confinement scale, and hence the glueball masses and the square root of the string tension, are proportional to $m_q^{2.18}$ at small quark masses.
While the approximation \eqref{approxLambda0} is based on 1-loop running and an abrupt mass threshold, it should become accurate in the limit $m_q/\Lambda_\text{UV} \rightarrow 0$ because the coupling near $\mu=m_q$ will be small, and the small coupling region dominates the evolution of the scale.

On the other hand, because $m_q/\Lambda_\text{IR} \gsim 1$ and grows as $m_q$ decreases, the 2-quark hadrons are effectively  ``heavy quark'' systems, with masses
\be
 m_{\text{Hadron}} \approx 2 m_q.
\ee
The size of the heavy quark system is proportional to its ``Bohr radius'', $[m_q g^2(m_q)]^{-1}$.

\begin{figure}[tbp]
\begin{centering}
\includegraphics[height=0.8\linewidth,keepaspectratio]{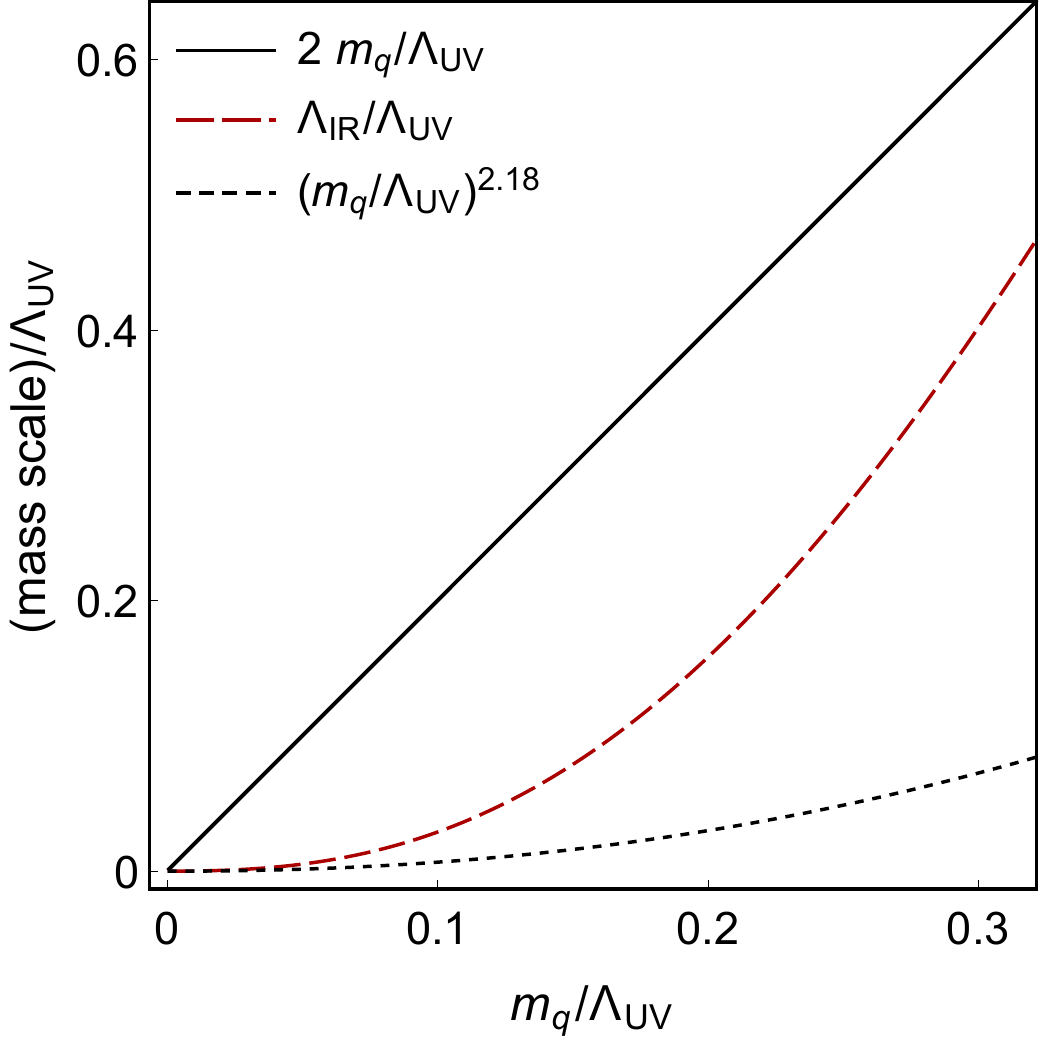}
\end{centering}
\caption{The expected scaling of meson/baryon masses $m_\text{Hadron} \approx 2m_q$ and the confinement scale $\Lambda_\text{IR}$, which is proportional to the glueball masses and square root of the string tension, as functions of quark mass $m_q$. The long-dashed, red line is obtained using the 2-loop massive quark $\beta$-function \eqref{eq:massivergeq} and the short-dashed, black line is the approximation \eqref{approxLambda0}.}
\label{fig:scaling}
\end{figure}

A more accurate estimation of the confinement scale can be obtained with the massive 2-loop $\beta$-function \cite{Jegerlehner:1998zg}.  In this case the mass threshold is automatically taken care by the $\beta$ and $\gamma$ -functions:
\begin{subequations}
\begin{align}
\mu \frac{d g^2}{d \mu}& = \beta\ssof{g^2,m_q/\mu}\ ,\\
\frac{\mu}{m_q} \frac{d m_q}{d \mu}& = -\gamma\ssof{g^2,m_q/\mu}\ .\label{eq:massivergeq}
\end{align}
\end{subequations}
Because also quark mass is running, we set the physical quark mass
at the initial scale $\mu_0 = 2 m_{q,0}$, give a range of initial values to $g^2(\mu_0)$ and evolve the equations to UV and IR until the coupling diverges.  The result $\Lambda_\text{IR}/\Lambda_\text{UV}$ is shown in Fig.~\ref{fig:scaling} as functions of $m_{q,0}/\Lambda_\text{UV}$.  We observe that this result agrees with the approximation \eqref{approxLambda0} at small $m_q$, but deviates from it substantially at larger $m_q$ when the mass threshold effects and higher order corrections affect the result significantly.  Our goal is to observe how well this behaviour is reproduced on the lattice.

\section{Lattice formulation}\label{sec:latform}
On the lattice the theory is defined by the action
\begin{displaymath}
S = S_G(U) + S_F(V) + c_{\rm SW}\,S_{\rm SW}(V)\ ,\label{lataction}
\end{displaymath}%
where $U$ is a $\SU{2}$ gauge link matrix in the fundamental representation, $V$ is a corresponding
smeared gauge link, defined by hypercubic truncated stout smearing
(HEX smearing)~\cite{Capitani:2006ni},
$S_G$ is the Wilson gauge action, and $S_F$ and $S_{\rm SW}$ are, respectively,
the Wilson fermion action and the clover term.  The parameters of the action are the bare lattice
gauge coupling $\beta_L \equiv 4/g_0^2$, appearing in $S_G$, the ``hopping parameter'' $\kappa$ in $S_F$,
and the Sheikholeslami-Wohlert coefficient $c_{\rm SW}$.  We set $c_{SW} = 1$, as is often used for HEX smeared fermions \cite{Rantaharju:2015yva}.

We simulate $N_{f}=24$ massive but mass-degenerate Wilson flavors on hyper-cubic,
toroidal lattices of
sizes $V=N_{s}^3\times N_{t}$, where $N_{s}$ and $N_{t}$ refer to the number of lattice sites
in spatial and temporal direction. These cover the values $N_{t}\in\cof{32,40,48}$ and
$N_{s}\in\cof{N_{t}/2,3\,N_{t}/4,N_{t}}$. The boundary conditions for the gauge field are
periodic in all directions, whereas for the
fermion fields, they are periodic only in spatial but anti-periodic in temporal direction,
as usual.

Simulations are carried out using a hybrid Monte Carlo (HMC) algorithm with leapfrog integrator
and chronological initial values for the fermion matrix inversion~\cite{Brower:1995vx}.
The HMC trajectories have unit-length and the number of leapfrog steps is tuned to yield
acceptance rates above 80\% (with a few exceptions).

The physical quark mass is determined by the lattice PCAC relation
\be
a\,m_q(x_4)=\frac{(\partial_4^\ast+\partial_4)f_A(x_4)}{4\,f_P(x_4)}\ ,
\ee
where $\partial_4$ and $\partial_4^{*}$ are, respectively, forward and backward lattice time-derivative operators, and
\iffalse
\beq
f_A(x_4) &=& -\frac{1}{L^3\,N_t}\sum\limits_{y,a}\avof{{}^{(w)}\!A^a_{\mu}\of{y_4}\,A^a_{\mu}\of{\mathbf{y},x_4+y_4}}\ \\
f_P(x_4) &=& \frac{1}{L^3\,N_t}\sum\limits_{y,a}\avof{{}^{(w)}\!P^a\of{y_4}\,P^a\of{\mathbf{y},x_4+y_4}}\
\eeq
\else
\beq
f_A(x_4) &=& -\frac{1}{N_s^3\,N_t}\sum\limits_{\mathbf{x},y,a}\avof{\!A^a_{\mu}\of{\mathbf{y},y_4}\,A^a_{\mu}\of{\mathbf{x},x_4+y_4}}\ ,\nonumber\\
f_P(x_4) &=& \frac{1}{N_s^3\,N_t}\sum\limits_{\mathbf{x},y,a}\avof{\!P^a\of{\mathbf{y},y_4}\,P^a\of{\mathbf{x},x_4+y_4}}\nonumber
\eeq
\fi
are the axial current and pseudoscalar density correlation functions
with point sources:
\beq
A^a_{\mu}\of{x}&=&\bar{\psi}\of{x}\gamma_{\mu} \gamma_5 \frac{1}{2} \tau^a \psi\of{x}\ ,\nonumber\\
P^a\of{x}&=&\bar{\psi}\of{x}\gamma_5 \frac{1}{2} \tau^a \psi\of{x}\ .\nonumber
\eeq
For each value of the bare gauge coupling $\beta_L\equiv 4/g_0^2$ the hopping parameter
$\kappa$ in the Wilson fermion action is tuned to cover a range of PCAC fermion masses
for which we study the behavior of the spectrum from the chiral limit to heavy quarks.
The masses of color singlet meson states are then determined by fitting the time sliced
average correlation functions with Coulomb gauge fixed wall sources:
\be
{}^{(w)}\!\Gamma^a\of{x_4}=\sum\limits_{\mathbf{x},\mathbf{y}}\bar{\psi}\of{\mathbf{x},x_4}\Gamma \frac{1}{2} \tau^a \psi\of{\mathbf{y},x_4}\ ,\nonumber
\ee
with $\Gamma$ representing an element of the Dirac algebra.

\section{Results}\label{sec:results}

Let us now turn to the details of the simulations and the results we have obtained on the spectrum of physical states.
The simulation parameters and corrsponding PCAC quark masses,
pseudoscalar ("$\pi$") and vector ("$\rho$") meson masses together with the acceptance rates
and accumulated statistic are given in Appendix~\ref{app:tables} in
tables~\ref{tbl:simparamnt24}--\ref{tbl:simparamnt48ns48}.

\paragraph*{Phase diagram:}

For the bare gauge coupling we use values
$\beta_L = 4/g_0^2 \in\cof{-0.25,0.001,0.25}$.
The small values of $\beta_{L}$ might at first seem to imply that the gauge field is deep in an unphysical
strong coupling ``bulk phase''. This is, however, in general not the case, as the Wilson fermions induce an
effective positive shift in $\beta_{L}$ \cite{Hasenfratz:1993az,Blum:1994xb}.
This is to leading order proportional to
the number of flavors and therefore substantial for $N_f=24$. Indeed, the above values of $\beta_L$ were successfully used in the measurement of the coupling constant evolution at vanishing fermion masses~\cite{Leino:2019qwk}.  A similar effect has been reported
in~\cite{deForcrand:2012vh} for staggered fermions.

The value of the $\beta_L$ affects the lattice spacing, but to the opposite direction to lattice QCD: lattice spacing is made smaller when $\beta_L$ is decreased.  Because of the Landau pole the theory on the lattice does not have a continuum limit.

\begin{figure*}[htbp]
\centering
\hfill
\begin{minipage}[t]{0.36\linewidth}
\centering
\includegraphics[height=\linewidth,keepaspectratio,right]{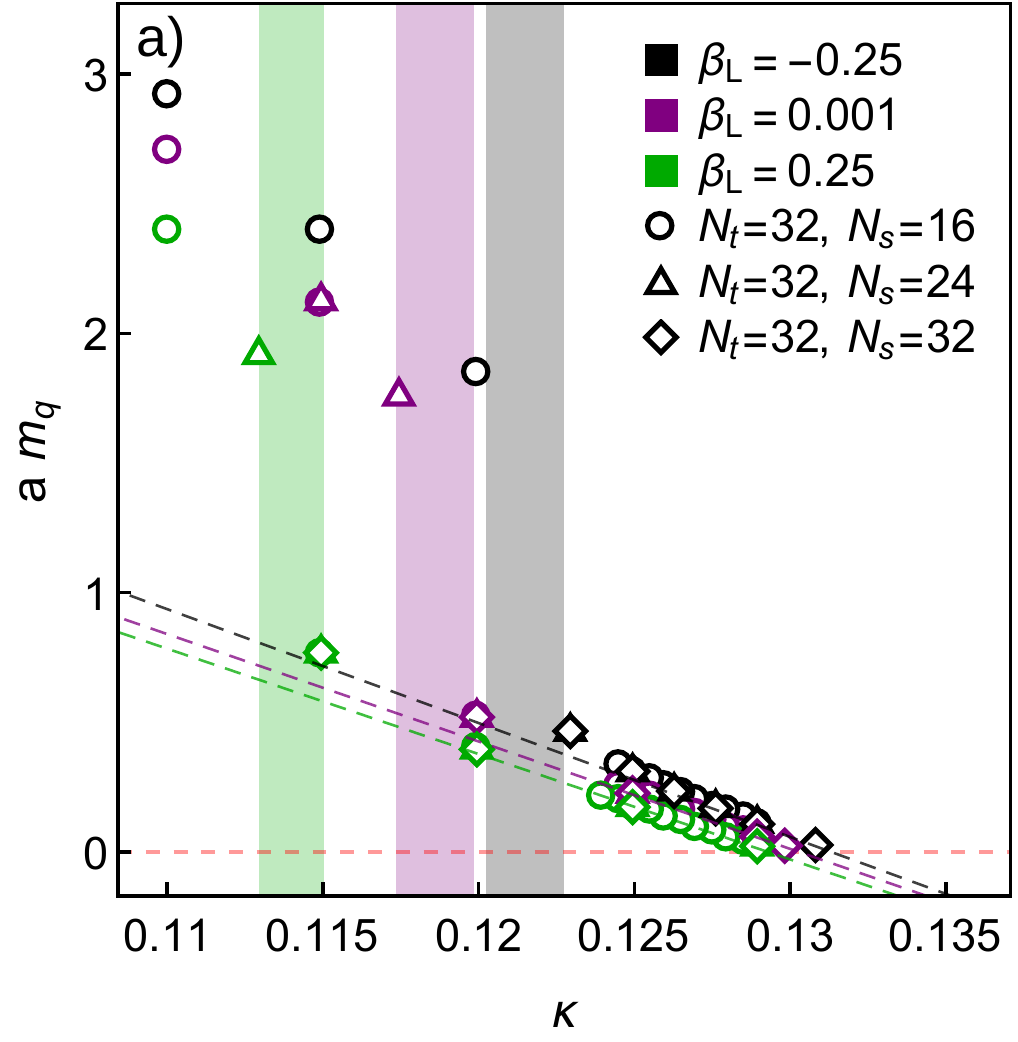}
\end{minipage}\hfill
\begin{minipage}[t]{0.36\linewidth}
\centering
\includegraphics[height=\linewidth,keepaspectratio,right]{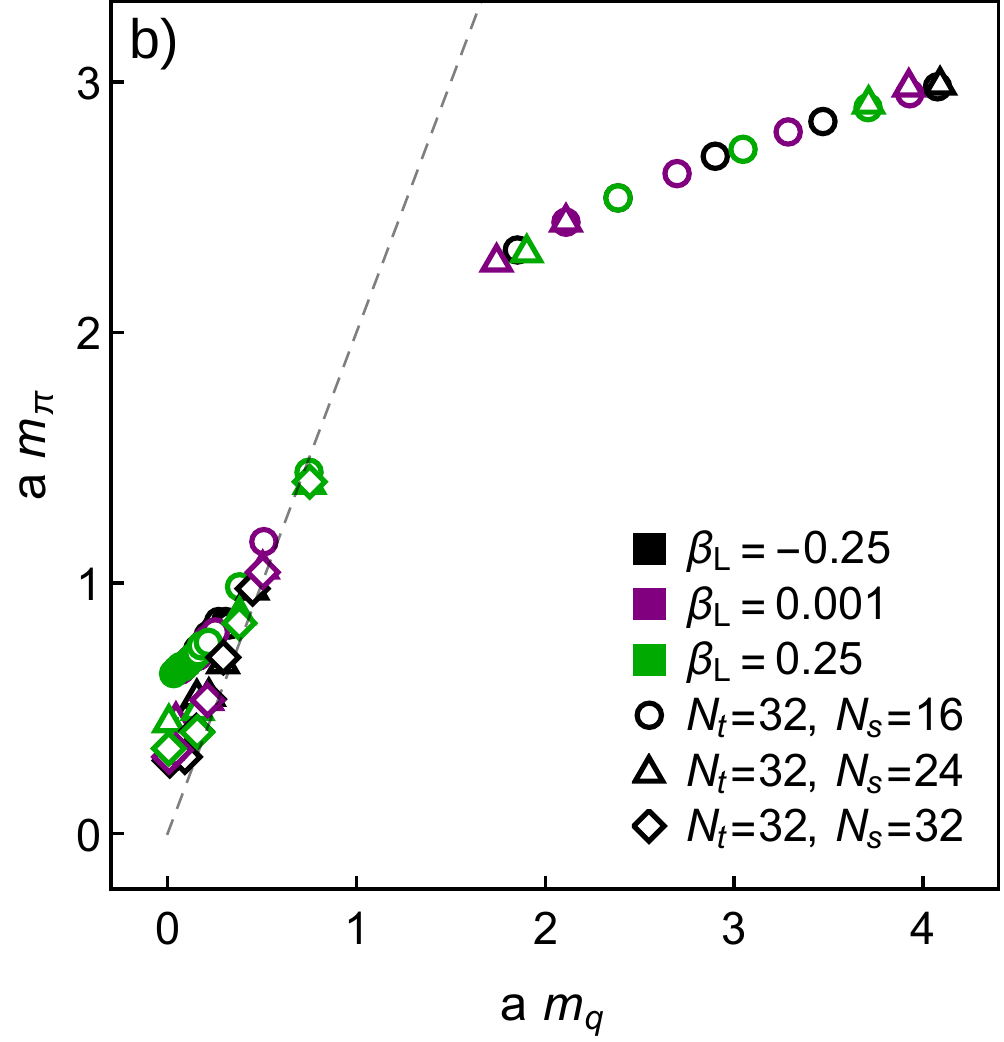}
\end{minipage}\hfill~\\[8pt]
\hfill
\begin{minipage}[t]{0.36\linewidth}
\centering
\includegraphics[height=\linewidth,keepaspectratio,right]{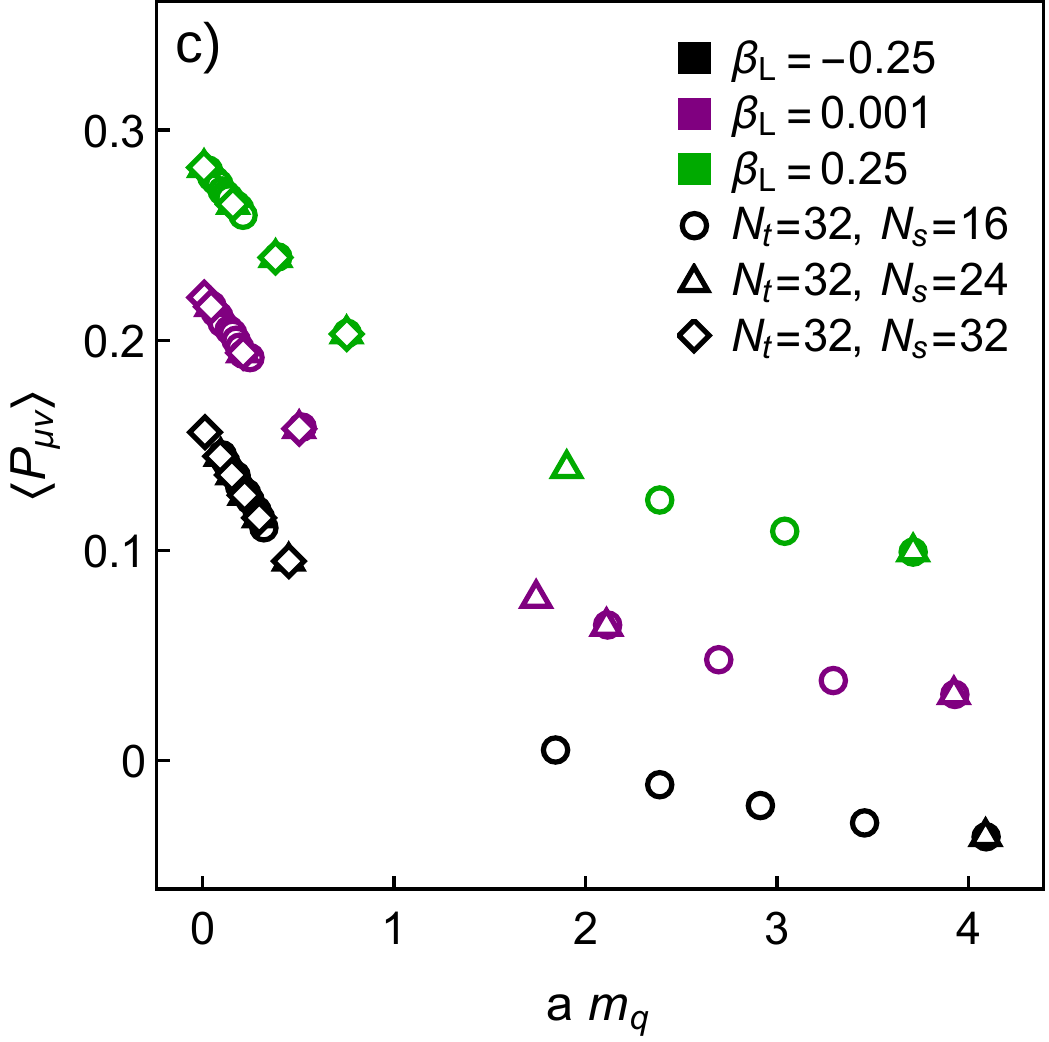}
\end{minipage}\hfill
\begin{minipage}[t]{0.36\linewidth}
\centering
\includegraphics[height=\linewidth,keepaspectratio,right]{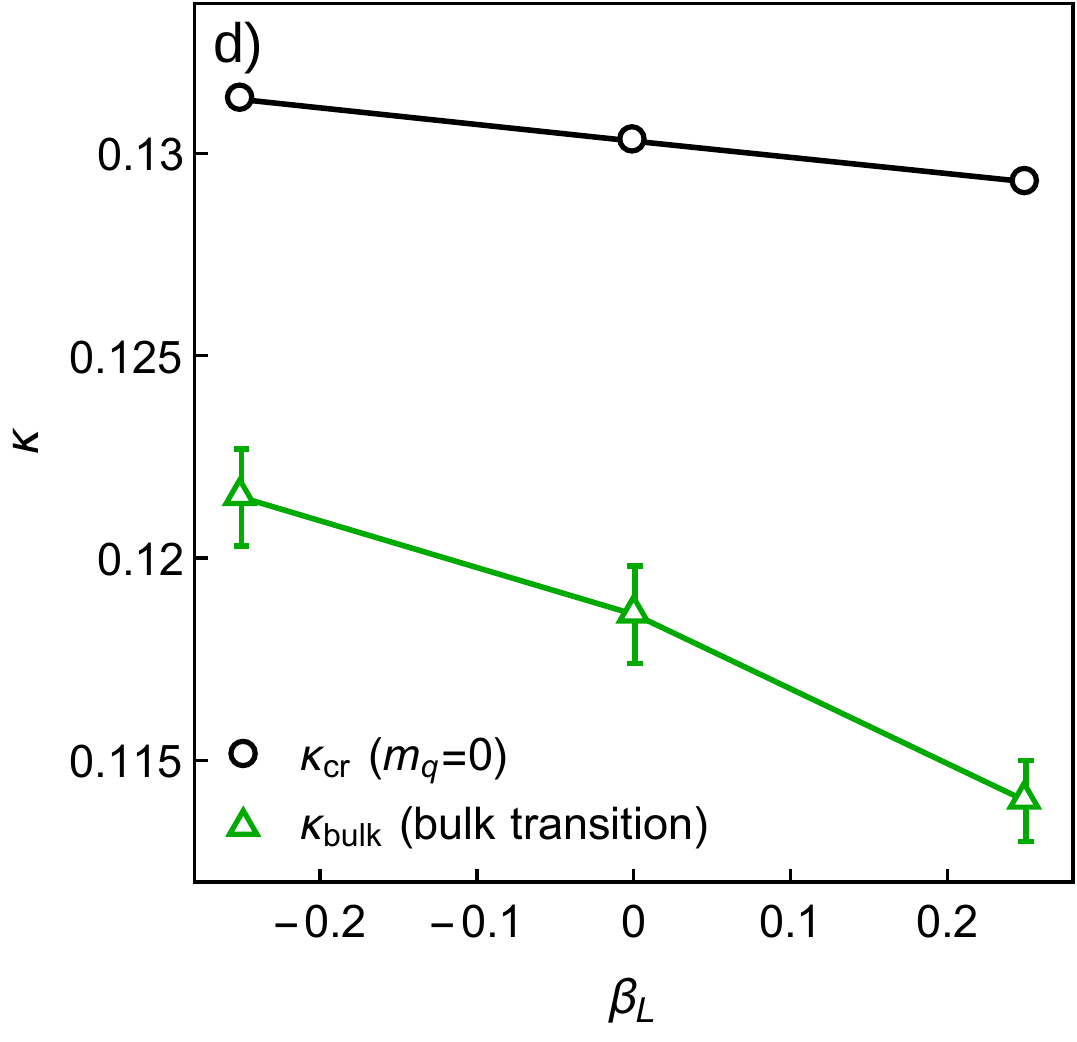}
\end{minipage}\hfill~
\caption{(a): The PCAC quark masses $m_q$ as functions of $\kappa$, measured at
$\beta_{L}=-0.25,\,0.001,\,0.25$ and lattice sizes $V=N_{s}^3\times N_{t}$ with $N_{t}=32$, $N_{s}=16,24,32$.
Here, and in what follows, different colours indicate the $\beta_L$-values and plot symbol shapes the volume.
The vertical shaded bands show the approximate locations of the bulk phase transitions at different $\beta_L$.
The dashed lines are linear fits to three smallest $m_q$-values for each $\beta_L$, used to determine the critical value $\kappa_{\text{cr}}$.
(b): The pseudoscalar meson (``pion'') mass $m_\pi$ and (c): plaquette as functions of $m_q$.  In (b) the dashed line is $m_\pi = 2 m_q$ -line.
(d): The phase diagram of the system.  The physical domain is between the bulk transition and critical ($m_q=0$) lines.}
\label{fig:bulkdemo}
\end{figure*}

In Fig.~\ref{fig:bulkdemo}(a) we show the measured PCAC quark masses $m_q$ as functions  of the hopping parameter $\kappa$.
It is evident that the system has an abrupt phase transition into unphysical ``bulk phase'' at small values of $\kappa$.  The approximate location of the bulk transition is shown with vertical shaded bands for each value of $\beta_L$. On the large-$\kappa$ side of the transition there is a range of $\kappa$ -values where $m_q$ decreases from 0.8--0.9 down to 0 as $\kappa$ is increased (here, and in what follows, dimensionful quantities are given in units of the lattice spacing $a$, unless specified otherwise).  This is the range where physics can be extracted.
The bulk transition is also visible in other observables, for example the pseudoscalar mass (Fig.~\ref{fig:bulkdemo}(b)) and plaquette expectation value (Fig.~\ref{fig:bulkdemo}(c)).  As expected, the pseudoscalar mass is close to $2m_q$.

The $(\beta_L,\kappa)$-plane phase diagram is shown in Fig.~\ref{fig:bulkdemo}(d).  The physically relevant domain is between the bulk transition and the critical line $\kappa_{\text{cr}}(\beta_L)$, where $m_q=0$.  The values of the critical $\kappa_{\text{cr}}(\beta_L)$ have been obtained by linear extrapolation of the measured $m_q(\kappa)$ at smallest quark masses.  The numerical values of $\kappa_{\text{cr}}$ and the bulk transition $\kappa_{\text{bulk}}$ are shown in Table~\ref{tbl:kappacrit}.

\begin{table}[tbp]
\centering
\begin{tabular}{c | c | c | c | c}
 $N_t$ & $N_s$ & $\beta_L$  & $\kappa_{\text{cr}}$ & $\kappa _{\text{bulk}}$ \\\hline
 32 & 32 & -0.25 & \text{0.131327(4)} & \text{0.1215(12)} \\
 32 & 32 & 0.001 & \text{0.130296(2)} & \text{0.1186(12)} \\
 32 & 32 & 0.25 & \text{0.129297(2)} & \text{0.1140(10)} \\
\end{tabular}
\caption{The table lists for a system of size $V=N_{s}^3\times N_{t}$ with $N_{s}=N_{t}=32$ and different $\beta$-values the values of $\kappa_{cr}$ and $\kappa_{bulk}$, being, respectively, the values of $\kappa$ for which the PCAC quark mass $m_{q}$ vanishes, and at which the bulk transition occurs.}
\label{tbl:kappacrit}
\end{table}

\paragraph*{Finite size effects:}

\begin{figure*}[htbp]
\centering
\hfill
\begin{minipage}[t]{0.36\linewidth}
\centering
\includegraphics[height=\linewidth,keepaspectratio]{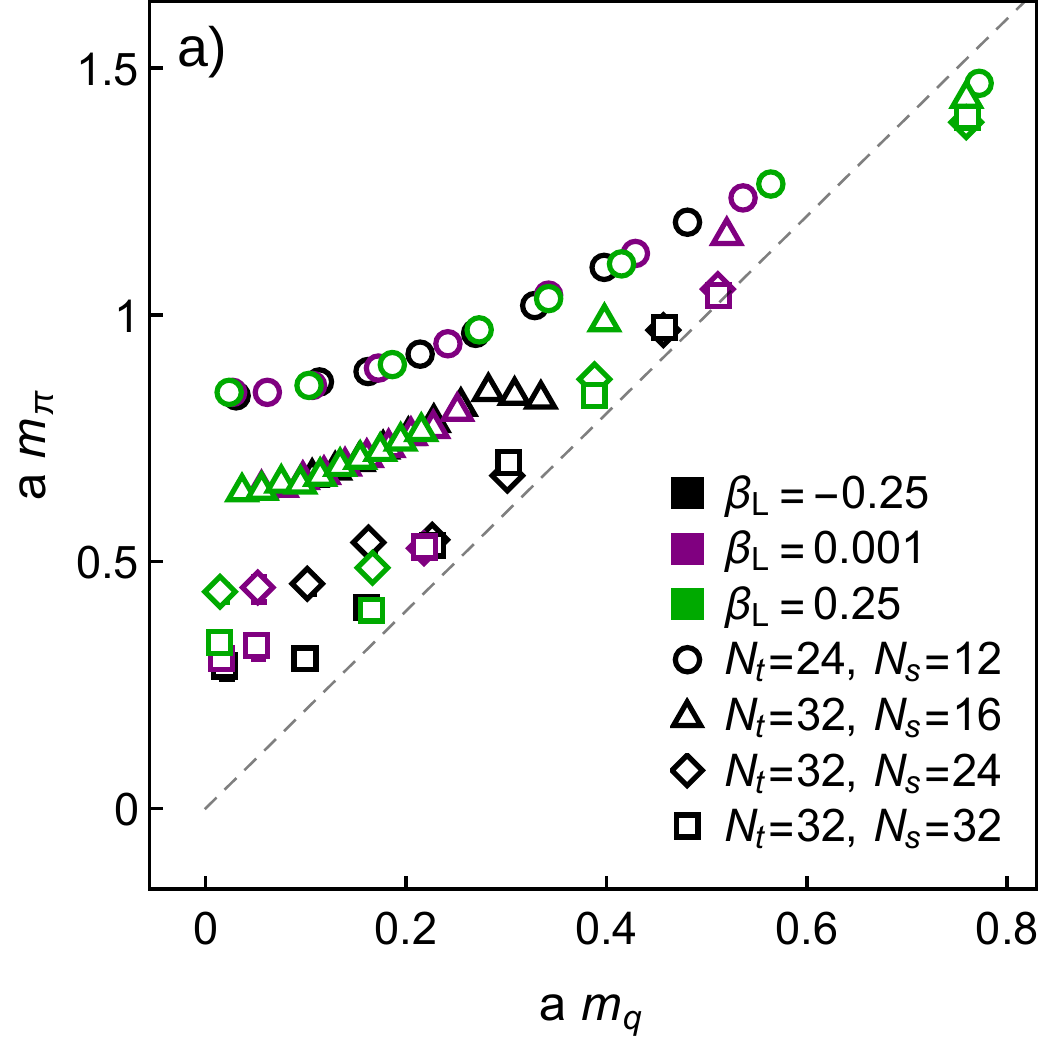}
\end{minipage}\hfill
\begin{minipage}[t]{0.36\linewidth}
\centering
\includegraphics[height=\linewidth,keepaspectratio]{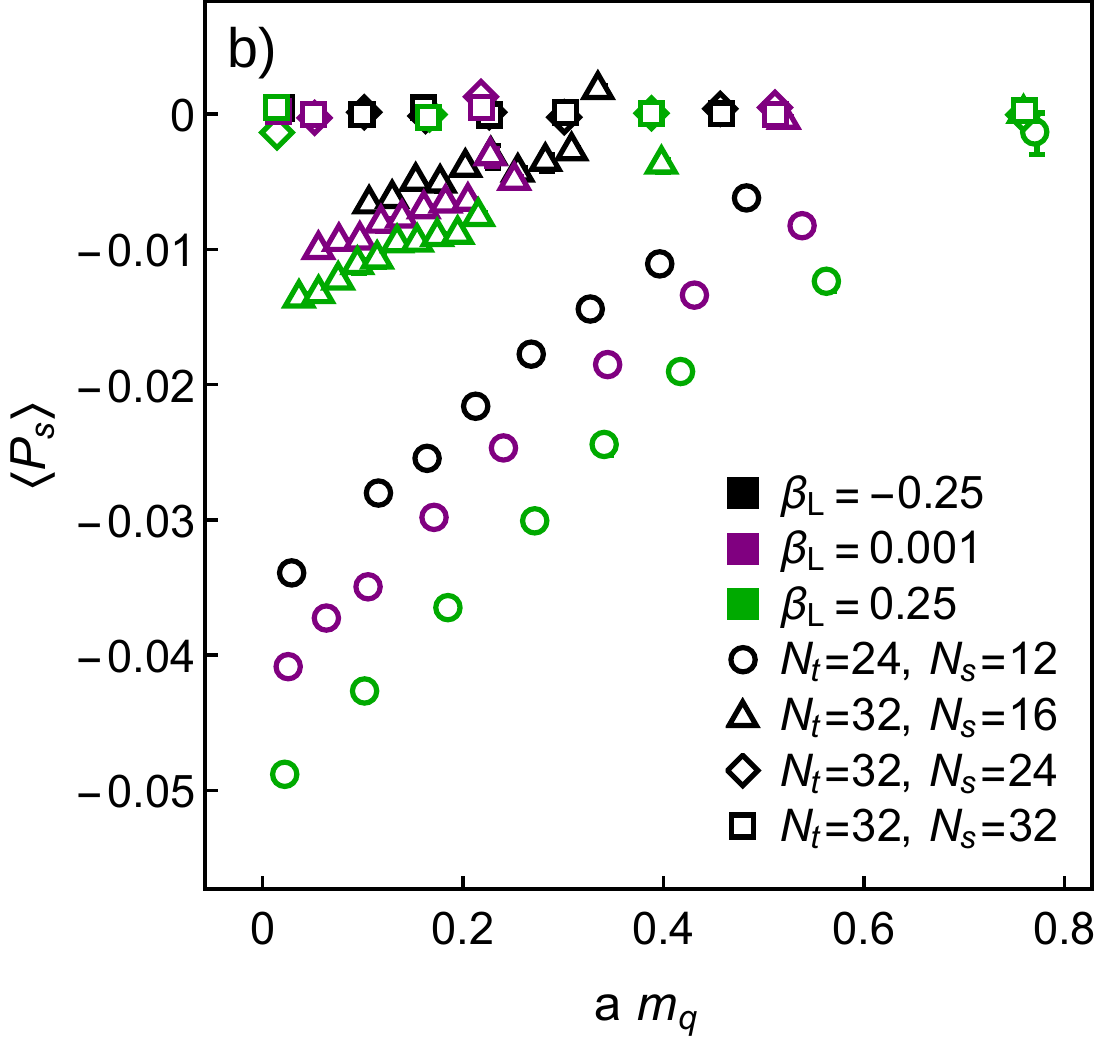}
\end{minipage}\hfill~
\caption{The figure shows for three different values of $\beta_{L}=-0.25,\,0.001,\,0.25$ and lattice sizes $V=N_{s}^3\times N_{t}$ with $N_{s}=16,24,32$ at $N_{t}=32$ and $N_{s}=12$ at $N_{t}=24$ (a) the pion mass $m_{\pi}$ and (b) the expectation value of spatial Polyakov loop $\avof{P_{s}}$ as functions of the PCAC quark mass $m_{q}$. Note that colors (black,purple,green) are used to distinguish between different values of $\beta_{L}$ and symbols (circle,triangle,diamond) are used to distinguish the different system sizes.}
\label{fig:finitevoleffects}
\end{figure*}

As discussed in Section~\ref{sec:spectrum}, the size of the hadrons grow as $m_q \rightarrow 0$, and we can expect large finite size effects in this limit.
This is evident in Fig.~\ref{fig:finitevoleffects}(a), where we show the pseudoscalar ``pion'' masses $m_\pi$ measured from volumes $N_s^3$ from $12^3$ up to $32^3$ at different values of $\beta_L$.  At small volumes the mass plateaus roughly at $m_\pi \approx 1/(N_s a)$ as $m_q$ is lowered.  Interestingly, there is very little dependence on the value of the lattice gauge coupling $\beta_L$, indicating that it is the quark mass which is important for finite size effects.

\begin{figure*}[htbp]
\hfill
\begin{minipage}[t]{0.36\linewidth}
\centering
\includegraphics[height=\linewidth,keepaspectratio]{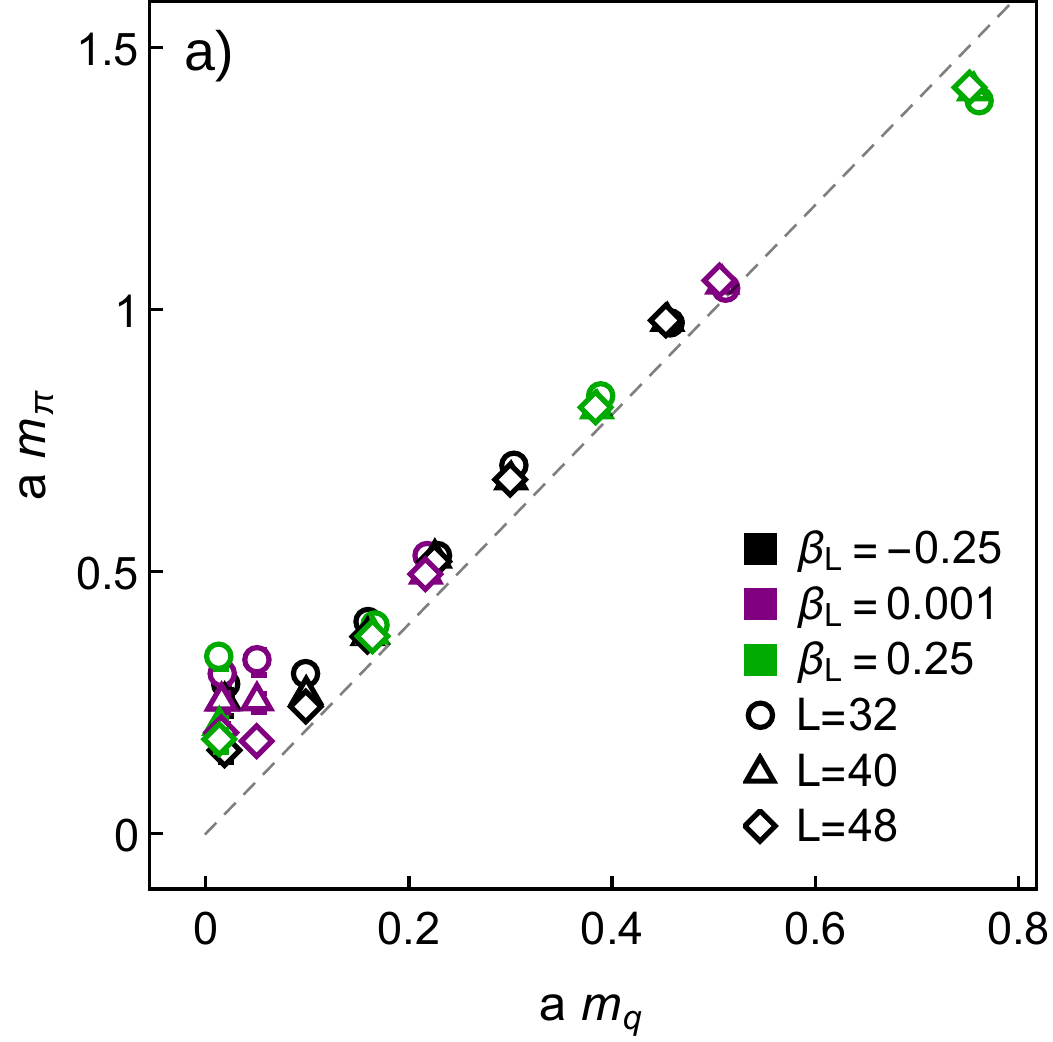}\\[8pt]
\includegraphics[height=\linewidth,keepaspectratio]{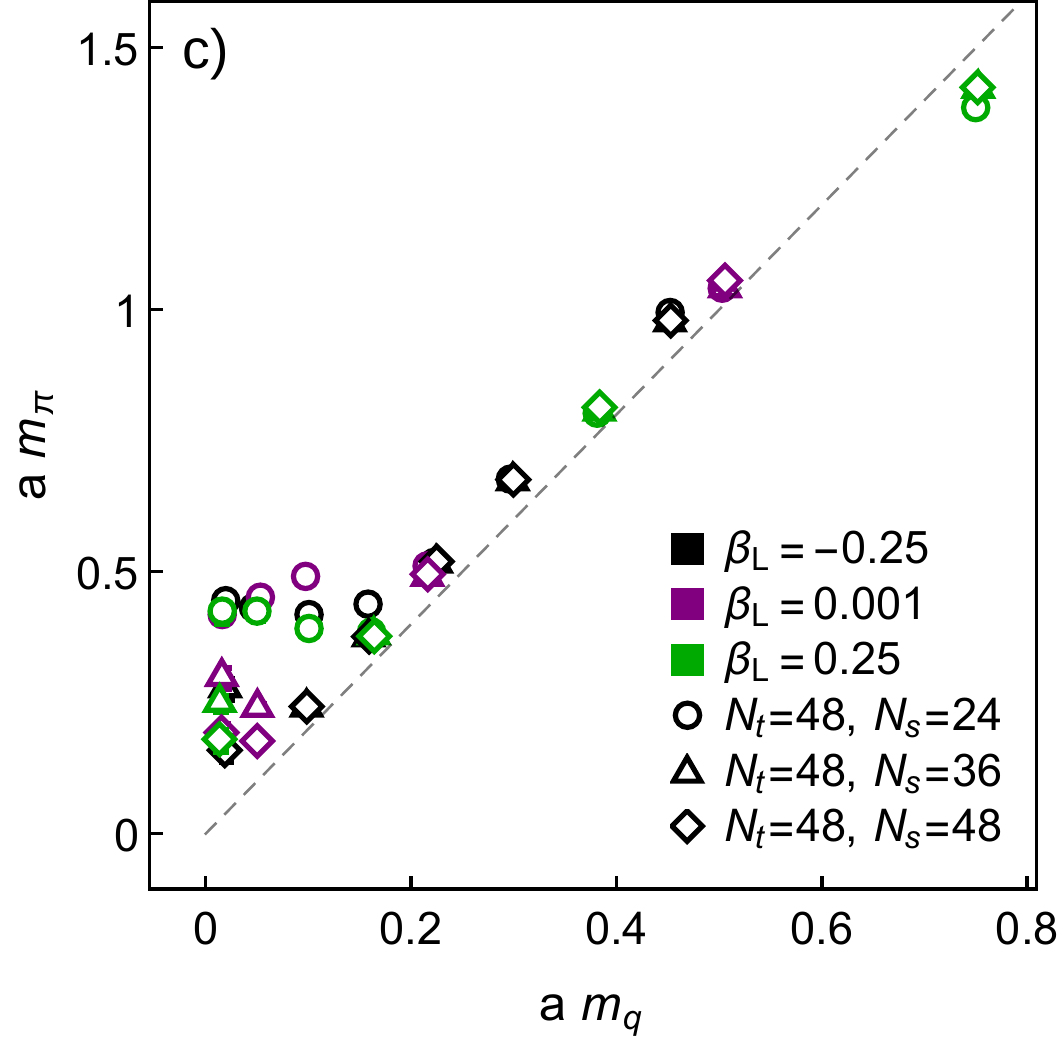}
\end{minipage}\hfill
\begin{minipage}[t]{0.36\linewidth}
\centering
\includegraphics[height=\linewidth,keepaspectratio]{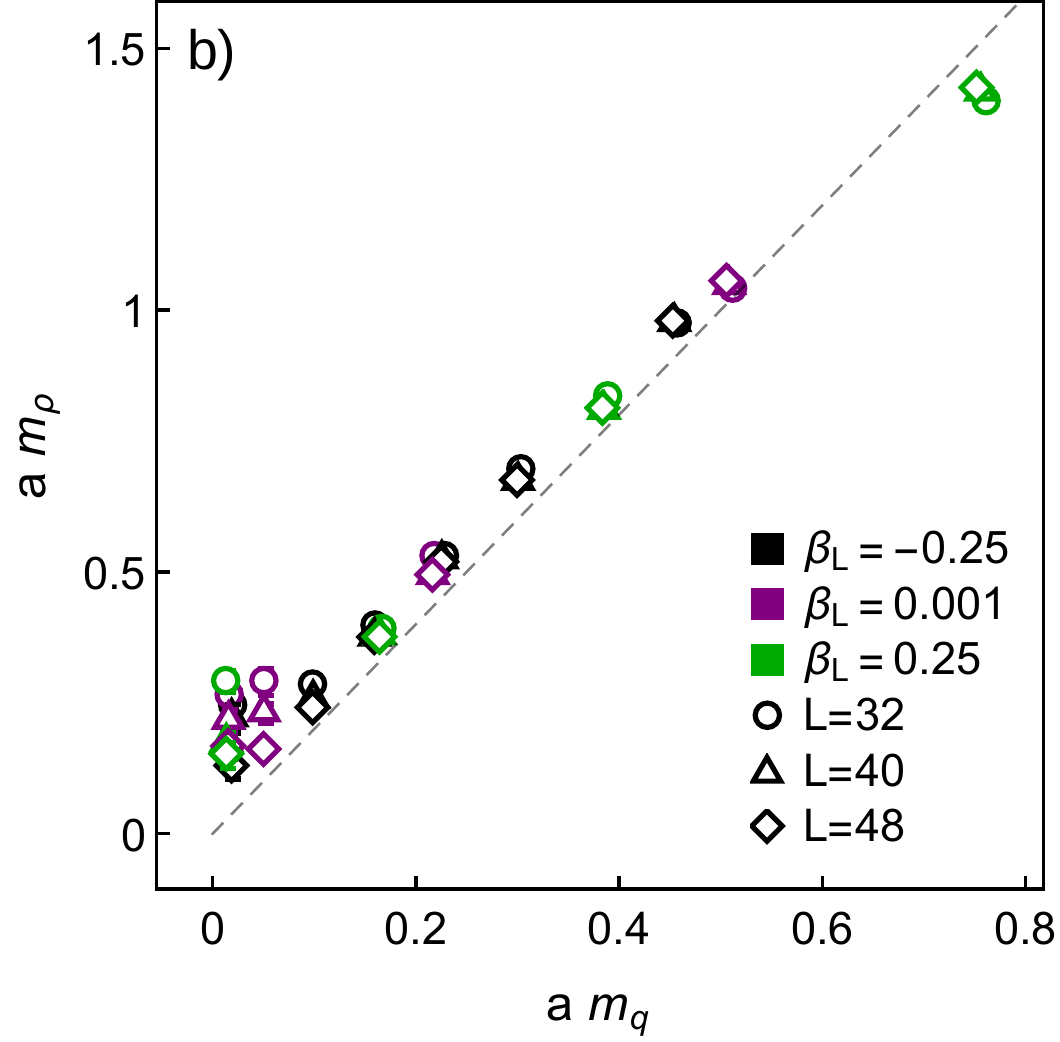}\\[8pt]
\includegraphics[height=\linewidth,keepaspectratio]{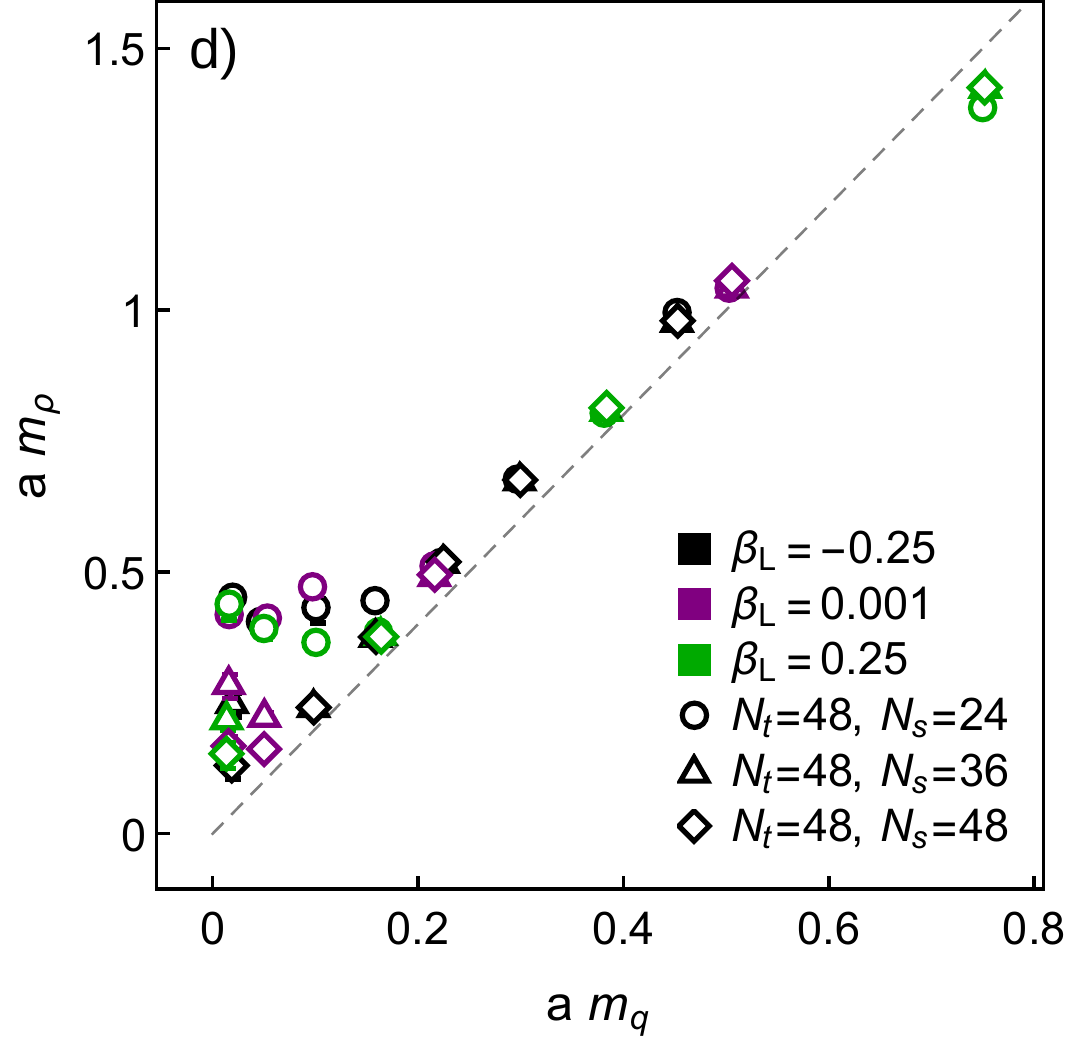}
\end{minipage}\hfill~
\caption{The pseudoscalar meson mass $m_\pi$ (a,c) and the vector meson mass $m_\rho$ (b,d) as functions of the PCAC quark mass $m_{q}$ for three values of $\beta_{L}=-0.25,\,0.001,\,0.25$ and system sizes $V=N_{s}^3\times N_{t}$, where in (a,b) $N_{t}=N_{s}=L$, $L=32,\,40,\,48$,
and in in (c,d) $N_{t}=48$, $N_{s}=24,\,36,\,48$. The colours (black,purple,green) are used to distinguish between different values of $\beta_{L}$ and symbols (circle,triangle,diamond) are used to distinguish the different system sizes.
}
\label{fig:mpivsmq}
\end{figure*}

In Fig.~\ref{fig:finitevoleffects}(b) we show the expectation values of the spatial Polyakov loop $\langle P \rangle$ as functions of the quark mass.  A non-zero value indicates that the system becomes spatially deconfined, i.e. the volume is too small to contain the confinement physics.  The negative expectation value is due to the periodic boundary conditions for fermions to spatial directions.
Clearly, at small volumes the magnitude of the expectation value of the Polyakov lines grow as the quark mass is decreased.  The magnitude of $\langle P \rangle$ is expected to decrease exponentially with the length of the line, $N_s$.  Nevertheless, the behaviour changes qualitatively as the volume is increased: $\langle P \rangle$
remains zero down to progressively smaller values of $m_q$.  At $N_s=32$ $\langle P \rangle$ remains zero at all $m_q$, within our statistical accuracy.

\paragraph*{Meson spectrum:}

The spectrum of pseudoscalar and vector mesons is shown in detail
in Fig.~\ref{fig:mpivsmq} for the range of $m_q$ away
from the bulk phase as discussed above. The pion masses as a function
of the quark mass are shown on panels a) and c), and the vector meson on panels b) and d).  The results corresponding to $\beta_L$
values $-0.25$, $0.001$ and $0.25$ are shown and different values are indicated by colors
black, blue and red respectively. In a) and b) the results are shown for system sizes
$N_t=N_s=L$ with $L=32$ (circles), $L=40$ triangles and $L=48$ (diamonds), while
in c) and d) the results are shown for $N_t=48$
with $N_s=24$ (circles), $N_s=36$ (triangles) and $N_s=48$ (diamonds).  Both the quark mass $m_q$ and the hadron masses are in units of the inverse lattice spacing.

As expected, the hadron masses very closely follow $2m_q$-line, slightly above it, and significant deviations appear only when $m_\text{hadron} \lsim 1/L$, the inverse spatial size of the lattice.  There is also no significant difference between psedoscalar and vector mesons within the accuracy of our measurements. This demonstrates the heavy quark nature of the mesons.

The results are independent of the bare lattice coupling $\beta_L$ in the sense that the results fall on a universal line on the $(m_q,m_\text{Hadron})$-plane.  We remind the reader that the lattice spacing becomes smaller as $\beta_L$ is decreased (bare coupling grows when lattice spacing is decreased), but the theory has no continuum limit because of the UV Landau pole.  Nevertheless, the observed universal behaviour in $\beta_L$ indicates that the theory has a scaling window where continuum-like physics can be studied.



\paragraph*{String tension and glueballs:}

In Section~\ref{sec:spectrum} it was predicted that the confinement scale, which determines the string tension and the glueball masses, scale as $(m_q/\Lambda_\text{UV})^{2.18}$ at small $m_q$, with deviations expected at larger $m_q$
(see Fig.~\ref{fig:scaling}).
We measure the
string tension $\sigma$ by constructing Wilson loops using spatially smeared gauge fields, with 5 different APE smearing levels up to 32 smearings \cite{Bolder:2000un}.  We measure Wilson loops
$W(\mathbf{r},t)$, where $\mathbf{r}$ is an integer multiple of one of the spatial vectors $(1,0,0)$, $(1,1,0)$, $(1,1,1)$, $(2,1,0)$,
$(2,1,1)$ + reflections and permutations.
The large-$t$ behaviour of the Wilson loops is fitted with an effective potential
\be
  \log \frac{W(\mathbf{r},t+1)}{W(\mathbf{r},t)} =
  \frac{A}{r} + \sigma\, r +  B,\label{eq:staticquarkpot}
\ee
where $r = |\mathbf{r}|$ and $A$, $B$ and $\sigma$ are fit parameters.

\begin{figure}[tbp]
\centering
\includegraphics[width=0.8\linewidth,keepaspectratio]{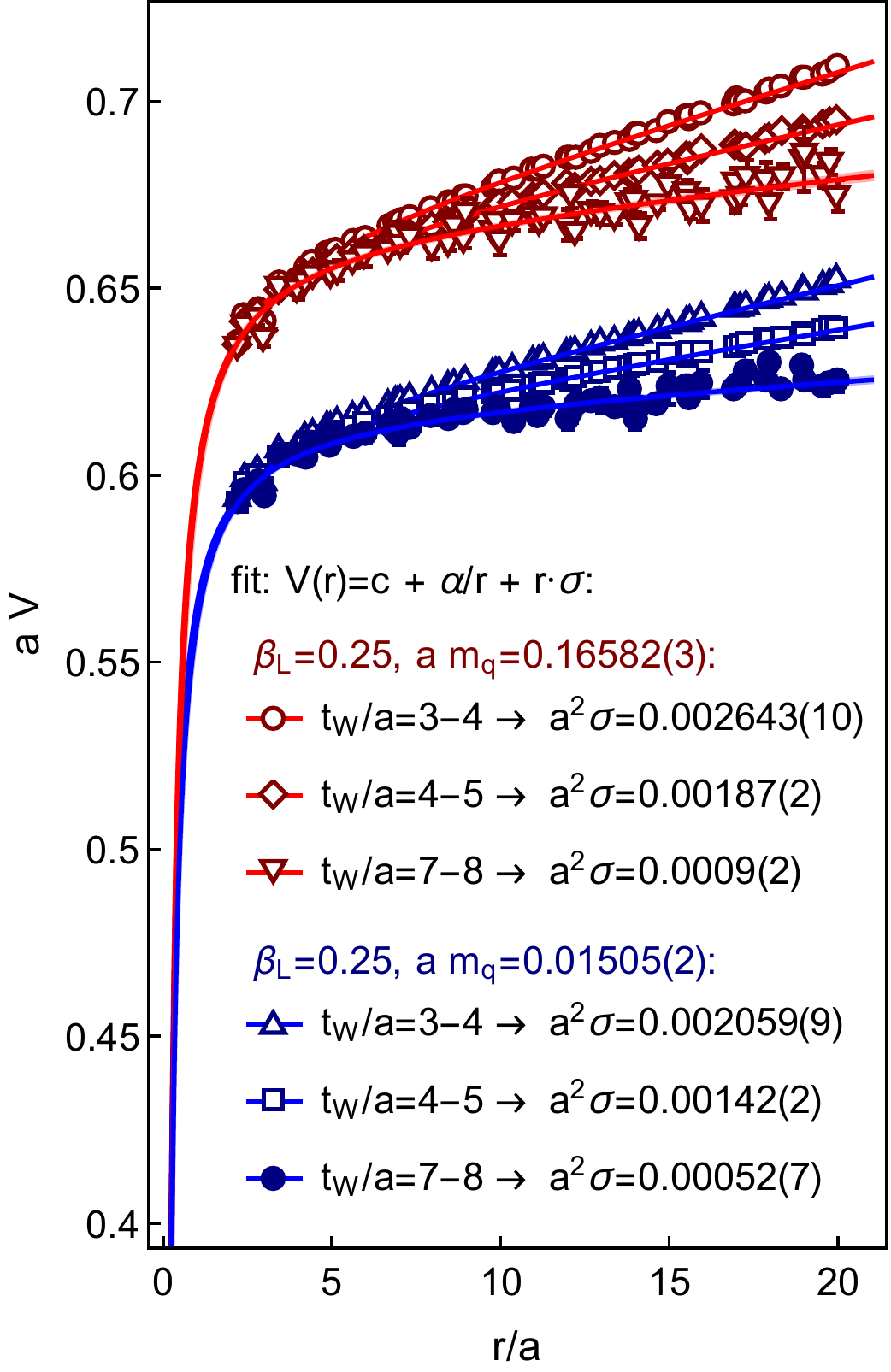}
\caption{Examples of fitting the effective potential while the $t$-distance of the Wilson loops, $t_w$, is varied.  The result does not settle to an asymptotic value, and the result is only an upper bound.}
\label{fig:potentialfit}
\end{figure}

However, it turns out that the string tension is too small in order to be reliably measured using our lattice sizes and statistics.  We can only present upper limits for the string tension, measured from the largest $t$-distance before the statistical errors make the measurement meaningless.  Nevertheless, the measurements have not yet stabilized as $t$ is changed.   This is shown in
Fig.~\ref{fig:potentialfit}.
We also attempted to do generalized eigenvalue analysis of the Wilson loops between different smearing levels, but this did not stabilize either, presumably because the spacing between eigenvalues is too small.

\begin{figure}[htbp]
\centering
\includegraphics[height=0.75\linewidth,keepaspectratio]{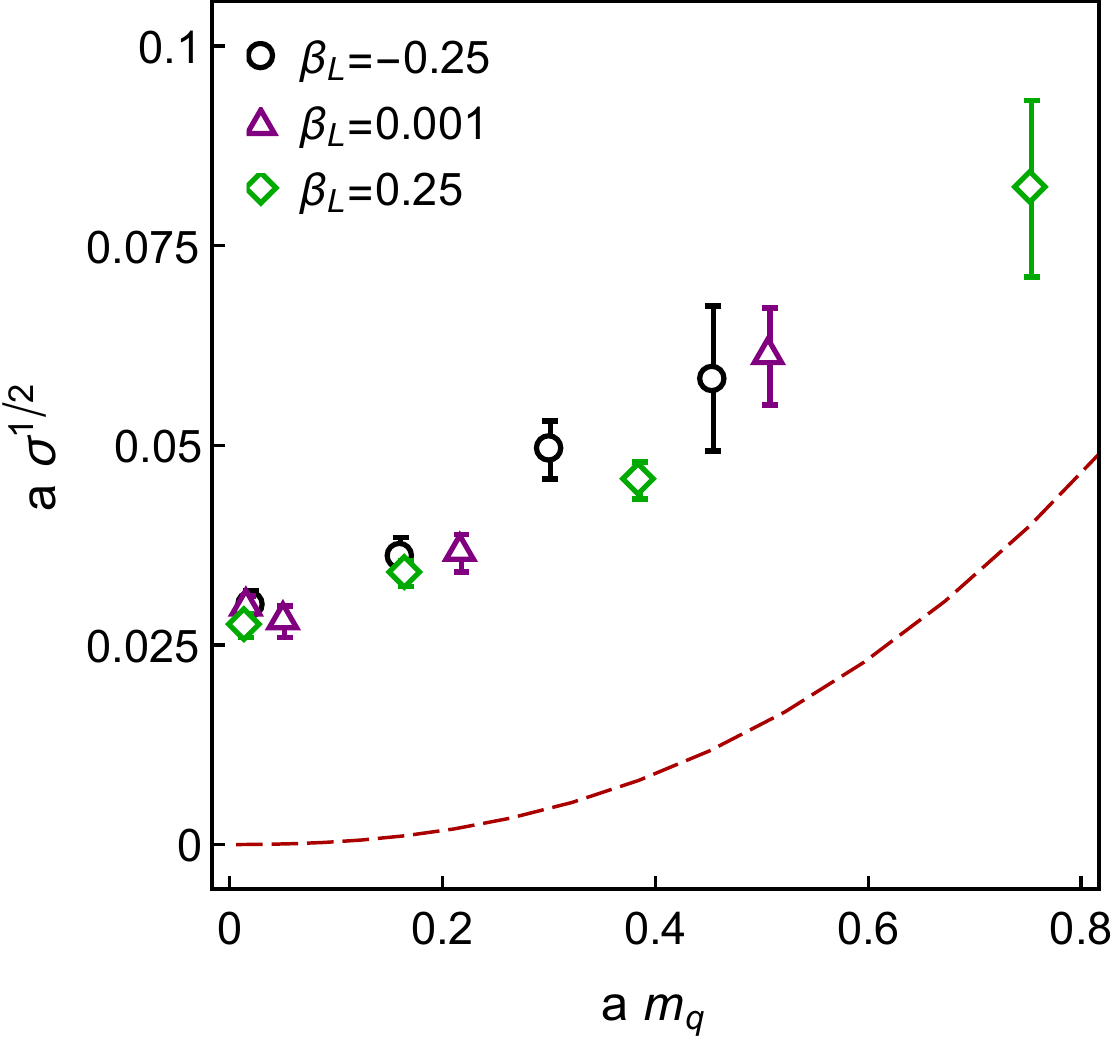}
\caption{The upper limit of the square root of the string tension $\sigma^{1/2}$ as function of the quark mass $m_q$.
The dashed line shows the confinement scale $\Lambda_{\mathrm{IR}}$ from Fig.~\ref{fig:scaling}, assuming (arbitrarily) that $a/\Lambda_{\mathrm{UV}}\sim 24$.
}
\label{fig:sigma}
\end{figure}

The upper limits of the square root of string tension are shown in Fig.~\ref{fig:sigma}.
To guide the expectation, the simple scaling ansatz \eqref{approxLambda0} is shown by the dashed line, with arbitrary scaling.

Finally, we attempt to measure glueball correlation functions using spatially smeared gauge links, with up to 100 APE smearing steps.  From the smeared gauge fields we construct operators coupling to scalar $J^{PC} = 0^{++}$ and tensor $2^{++}$ glueballs.  On the lattice the operators transform under cubic group representations, and we measure operators under 6 different representations.  Considering the difficulties in measuring the string tension, it is perhaps not surprising that we were not able to reliably measure glueball masses.  The correlation functions coupling to the $O^{++}$ state are very noisy due to the disconnected contribution.  Correlation functions of operators in the cubic group representations $E^{++}$ and $T_2^{++}$, which couple to the continuum $2^{++}$ state, do not have a disconnected part and are somewhat better behaved.  Nevertheless, we were not able to obtain the asymptotic state in these channels.

We assign the failure of the glueball mass measurement to the (expected) small value of the masses. This implies that the numerous excited states coupling to the same operators also have small masses, and it becomes very difficult to find the ground state.  We attempted to use generalized eigenvalue analysis of operators with different number of smearing steps, but the small mass gaps between states rendered the procedure unstable.

\begin{figure}[htbp]
\centering
\includegraphics[height=0.8\linewidth,keepaspectratio]{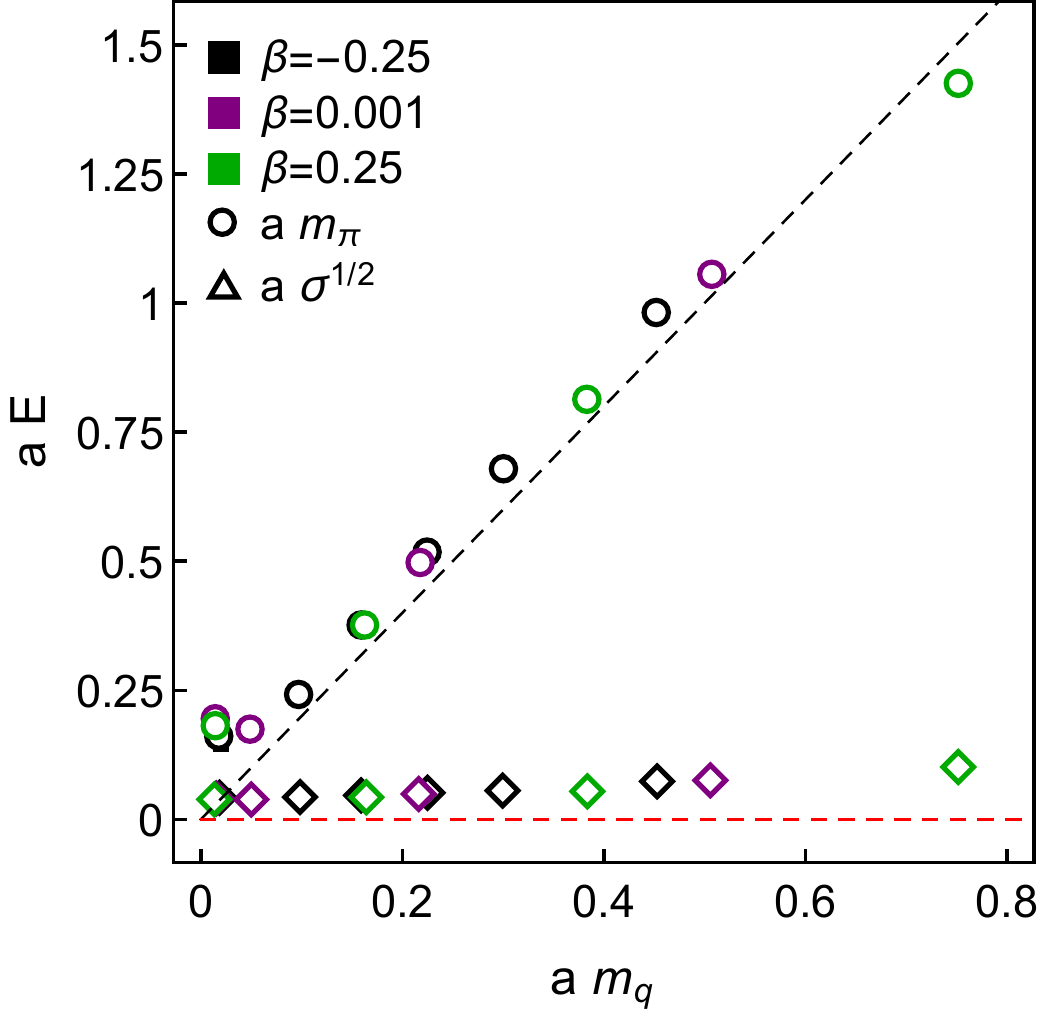}
\caption{The pseudoscalar mass $m_\pi$ and the upper limit of the square root of string tension $\sigma^{1/2}$ as functions of $m_q$.  The data is measured from lattices of size $48^4$.}
\label{fig:summary}
\end{figure}

In Fig.~\ref{fig:summary} we plot together the pseudoscalar mass and the upper limit of the square root of the string tension, measured from $48^4$ lattices.  $m_\pi$ is more than an order of magnitude larger than the upper limit of $\sigma^{1/2}$, and at the smaller end of the $m_\pi$-range the true hierarchy is expected to become much larger.  Comparing with Fig.~\ref{fig:scaling}, we can conclude that the physics on our lattices correspond to the case where quark masses are substantially below the Landau pole, $m_q/\Lambda_\text{UV} \ll 1$.  Because in the lattice units $m_q a \lsim 1$, this implies that the Landau pole is effectively at considerably larger scale than the inverse lattice spacing, $\Lambda_\text{UV} \gg 1/a$.  This was also observed in our earlier study of the running coupling at massless theory \cite{Leino:2019qwk}.

It is difficult to reach larger values of $m_q/\Lambda_\text{UV}$ with lattice simulations: trying to make the bare coupling stronger and the quarks heavier moves us towards the direction of strong lattice artifacts.  With our choice of the lattice action the bulk transition prevents us from using heavier quarks.

\section{Conclusions}\label{sec:conclusion}

We have studied the mass spectrum of SU(2) gauge theory with $N_f=24$ Dirac fermions on the lattice, complementing
our earlier analysis~\cite{Leino:2019qwk} on the running coupling in this theory.
The massless theory is free in the infrared, without bound states.  When the quark mass $m_q $ is non-zero the behaviour is different: in the infrared the theory is confining, and its spectrum includes mesons, two-quark baryons and glueballs, and the string tension is non-vanishing.  We have presented scaling relations for the hadron masses and for the confinement scale as functions of the quark mass.  The hadrons behave effectively as heavy quark systems with masses close to $2m_q$, whereas the confinement scale decreases faster, at small $m_q$ proportional to $m_q^{2.18}$.

The scaling relations imply that glueball masses and square root of the string tension $\sigma^{1/2}$ are much smaller than hadron masses unless quarks are very heavy, with masses of the order of the Landau pole of the theory.  The results from lattice simulations confirm this behaviour: $\sigma^{1/2}$ is more than an order of magnitude smaller than the hadron masses.  However, we were only able to give an upper limit for $\sigma^{1/2}$ in the range of quark masses studied, and we expect the true result to be substantially below the upper limit at small $m_q$.

To fully verify the behaviour of the confinement scale as $m_q$ is varied we would need to use much heavier quarks and stronger lattice coupling, so that the quark mass should be close to the Landau pole, $m_q/\Lambda_\text{UV} \lsim 1$. However, lattice bulk transition and increased lattice artifacts prevent us from moving significantly to that direction.  Furhermore, because the theory is only defined up to the UV cutoff scale, moving quark masses close to it makes the relevant physics sensitive to the details of the UV regularization used.

\section{Acknowledgement}
\label{Aknwldg}
The support of the Academy of Finland grants 308791, 310130, 319066 and 320123 is acknowledged. The authors wish to acknowledge CSC - IT Center for Science, Finland, for generous computational resources.

\appendix
\section{Tables of simulation parameters and results}
\label{app:tables}

A summary of the simulation parameters and corresponding PCAC quark masses, pion and rho-meson masses, as
well as the acceptance rates and accumulated statistics (after subtraction of non-thermalized
configurations), is given in the tables~\ref{tbl:simparamnt24}--\ref{tbl:simparamnt48ns48}.

\begin{table}[h]
\centering
\begin{tabular}{c | c | c | c | c | c | c | c | r}
$N_{t}$ & $N_{s}$ & $\beta_{L}$ & $\kappa$ & $m_q$ & $m_{\pi}$ & $m_{\rho}$ & acc. & stat. \\\hline
 24 & 12 & -0.25 & 0.1309 & \text{0.037(2)} & \text{0.84(5)} & \text{0.91(5)} & 0.98 & \text{8.8k} \\
 24 & 12 & -0.25 & 0.129 & \text{0.133(5)} & \text{0.79(5)} & \text{0.80(7)} & 0.98 & \text{9k} \\
 24 & 12 & -0.25 & 0.128 & \text{0.194(5)} & \text{0.86(3)} & \text{0.87(4)} & 0.99 & \text{9.3k} \\
 24 & 12 & -0.25 & 0.127 & \text{0.263(5)} & \text{0.91(3)} & \text{0.93(3)} & 0.99 & \text{9.5k} \\
 24 & 12 & -0.25 & 0.126 & \text{0.337(8)} & \text{0.98(4)} & \text{0.99(4)} & 0.99 & \text{9.7k} \\
 24 & 12 & -0.25 & 0.125 & \text{0.40(2)} & \text{1.00(4)} & \text{1.00(5)} & 0.99 & \text{9.7k} \\
 24 & 12 & -0.25 & 0.124 & \text{0.49(2)} & \text{1.08(3)} & \text{1.09(4)} & 0.99 & \text{9.5k} \\
 24 & 12 & -0.25 & 0.123 & \text{0.58(2)} & \text{1.13(3)} & \text{1.14(3)} & 0.99 & \text{9.8k} \\
 24 & 12 & 0.001 & 0.1299 & \text{0.029(1)} & \text{0.77(4)} & \text{0.76(6)} & 0.99 & \text{9.2k} \\
 24 & 12 & 0.001 & 0.129 & \text{0.077(3)} & \text{0.86(4)} & \text{0.88(5)} & 0.99 & \text{9.4k} \\
 24 & 12 & 0.001 & 0.128 & \text{0.129(3)} & \text{0.85(3)} & \text{0.84(4)} & 0.99 & \text{9.5k} \\
 24 & 12 & 0.001 & 0.1265 & \text{0.202(6)} & \text{0.85(4)} & \text{0.87(4)} & 0.99 & \text{9.5k} \\
 24 & 12 & 0.001 & 0.125 & \text{0.288(9)} & \text{0.93(4)} & \text{0.95(4)} & 0.99 & \text{9.6k} \\
 24 & 12 & 0.001 & 0.123 & \text{0.411(8)} & \text{1.03(3)} & \text{1.03(3)} & 0.99 & \text{10.2k} \\
 24 & 12 & 0.001 & 0.1215 & \text{0.50(2)} & \text{1.06(3)} & \text{1.07(3)} & 0.99 & \text{10.1k} \\
 24 & 12 & 0.001 & 0.12 & \text{0.66(1)} & \text{1.22(2)} & \text{1.22(2)} & 0.99 & \text{10.2k} \\
 24 & 12 & 0.25 & 0.129 & \text{0.0270(8)} & \text{0.81(4)} & \text{0.79(5)} & 0.99 & \text{9.6k} \\
 24 & 12 & 0.25 & 0.127 & \text{0.121(3)} & \text{0.82(3)} & \text{0.85(4)} & 0.99 & \text{9.9k} \\
 24 & 12 & 0.25 & 0.125 & \text{0.231(6)} & \text{0.93(3)} & \text{0.96(3)} & 0.99 & \text{10.1k} \\
 24 & 12 & 0.25 & 0.123 & \text{0.33(1)} & \text{0.98(4)} & \text{0.99(4)} & 0.99 & \text{10.6k} \\
 24 & 12 & 0.25 & 0.1215 & \text{0.392(9)} & \text{0.98(4)} & \text{1.00(4)} & 0.99 & \text{10.5k} \\
 24 & 12 & 0.25 & 0.12 & \text{0.50(1)} & \text{1.10(3)} & \text{1.11(3)} & 0.99 & \text{10.3k} \\
 24 & 12 & 0.25 & 0.1175 & \text{0.64(2)} & \text{1.16(4)} & \text{1.16(4)} & 0.99 & \text{6k} \\
 24 & 12 & 0.25 & 0.115 & \text{0.92(2)} & \text{1.41(2)} & \text{1.41(2)} & 0.99 & \text{11.5k} \\
\end{tabular}
\caption{The table shows for systems of size $V=N_{s}^3\times N_{t}$ with $N_{t}=24$ and $N_{s}=12$ the simulated values of $\beta_{L}$ and $\kappa$, the corresponding PCAC quark mass, $m_{q}$, pion and $\rho$-meson masses, $m_{\pi}$, $m_{\rho}$, the acceptance rate of the HMC trajectories, and the number of usable configurations (after thermalization).}
\label{tbl:simparamnt24}
\end{table}

\begin{table}[h]
\centering
\begin{tabular}{c | c | c | c | c | c | c | c | r}
$N_{t}$ & $N_{s}$ & $\beta_{L}$ & $\kappa$ & $m_q$ & $m_{\pi}$ & $m_{\rho}$ & acc. & stat. \\\hline
 32 & 32 & -0.25 & 0.1309 & \text{0.0207(2)} & \text{0.28(3)} & \text{0.24(3)} & 0.89 & \text{0.4k} \\
 32 & 32 & -0.25 & 0.129 & \text{0.1008(2)} & \text{0.299(5)} & \text{0.284(5)} & 0.84 & \text{0.5k} \\
 32 & 32 & -0.25 & 0.1277 & \text{0.1619(2)} & \text{0.403(6)} & \text{0.397(6)} & 0.84 & \text{1.2k} \\
 32 & 32 & -0.25 & 0.1263 & \text{0.2282(2)} & \text{0.530(2)} & \text{0.530(2)} & 0.84 & \text{1.3k} \\
 32 & 32 & -0.25 & 0.125 & \text{0.3043(2)} & \text{0.697(6)} & \text{0.695(7)} & 0.82 & \text{1.5k} \\
 32 & 32 & -0.25 & 0.123 & \text{0.4595(5)} & \text{0.971(3)} & \text{0.971(3)} & 0.84 & \text{1.7k} \\
 32 & 32 & 0.001 & 0.1299 & \text{0.0176(1)} & \text{0.30(2)} & \text{0.26(2)} & 0.89 & \text{1.4k} \\
 32 & 32 & 0.001 & 0.129 & \text{0.0526(1)} & \text{0.33(3)} & \text{0.29(3)} & 0.83 & \text{0.9k} \\
 32 & 32 & 0.001 & 0.125 & \text{0.2201(2)} & \text{0.53(2)} & \text{0.52(2)} & 0.83 & \text{1.4k} \\
 32 & 32 & 0.001 & 0.12 & \text{0.5136(5)} & \text{1.036(2)} & \text{1.036(2)} & 0.85 & \text{1.7k} \\
 32 & 32 & 0.25 & 0.129 & \text{0.0155(1)} & \text{0.33(2)} & \text{0.29(2)} & 0.86 & \text{0.7k} \\
 32 & 32 & 0.25 & 0.125 & \text{0.1669(1)} & \text{0.398(2)} & \text{0.391(2)} & 0.89 & \text{1.1k} \\
 32 & 32 & 0.25 & 0.12 & \text{0.3894(2)} & \text{0.833(5)} & \text{0.832(5)} & 0.88 & \text{1.5k} \\
 32 & 32 & 0.25 & 0.115 & \text{0.762(2)} & \text{1.40(2)} & \text{1.40(2)} & 0.88 & \text{1.9k} \\
\end{tabular}
\caption{Same as Table \ref{tbl:simparamnt24} for $N_t=32$, $N_s=32$.}
\label{tbl:simparamnt32}
\end{table}

\begin{table}[h]
\centering
\begin{tabular}{c | c | c | c | c | c | c | c | r}
$N_{t}$ & $N_{s}$ & $\beta_{L}$ & $\kappa$ & $m_q$ & $m_{\pi}$ & $m_{\rho}$ & acc. & stat. \\\hline
 40 & 20 & -0.25 & 0.1309 & \text{0.0226(1)} & \text{0.50(2)} & \text{0.50(2)} & 0.95 & \text{2k} \\
 40 & 20 & -0.25 & 0.1302 & \text{0.0509(1)} & \text{0.48(2)} & \text{0.51(2)} & 0.97 & \text{1.3k} \\
 40 & 20 & -0.25 & 0.129 & \text{0.1045(2)} & \text{0.56(1)} & \text{0.54(2)} & 0.86 & \text{1k} \\
 40 & 20 & -0.25 & 0.1277 & \text{0.1646(2)} & \text{0.51(2)} & \text{0.48(2)} & 0.85 & \text{0.6k} \\
 40 & 20 & -0.25 & 0.1263 & \text{0.2292(3)} & \text{0.64(2)} & \text{0.65(2)} & 0.88 & \text{0.3k} \\
 40 & 20 & -0.25 & 0.125 & \text{0.2978(5)} & \text{0.663(3)} & \text{0.663(3)} & 0.89 & \text{0.2k} \\
 40 & 20 & -0.25 & 0.123 & \text{0.4538(4)} & \text{0.971(4)} & \text{0.972(4)} & 0.86 & \text{2k} \\
 40 & 20 & 0.001 & 0.1299 & \text{0.0189(1)} & \text{0.53(2)} & \text{0.57(2)} & 0.97 & \text{2k} \\
 40 & 20 & 0.001 & 0.129 & \text{0.0547(1)} & \text{0.52(2)} & \text{0.57(2)} & 0.88 & \text{1k} \\
 40 & 20 & 0.001 & 0.1278 & \text{0.1019(1)} & \text{0.56(2)} & \text{0.54(2)} & 0.96 & \text{1.4k} \\
 40 & 20 & 0.001 & 0.125 & \text{0.2223(2)} & \text{0.61(2)} & \text{0.63(2)} & 0.91 & \text{0.5k} \\
 40 & 20 & 0.001 & 0.12 & \text{0.5099(5)} & \text{1.10(2)} & \text{1.10(2)} & 0.9 & \text{1.5k} \\
 40 & 20 & 0.25 & 0.129 & \text{0.0168(1)} & \text{0.48(2)} & \text{0.50(3)} & 0.88 & \text{1k} \\
 40 & 20 & 0.25 & 0.1281 & \text{0.0523(1)} & \text{0.51(2)} & \text{0.53(3)} & 0.97 & \text{1.4k} \\
 40 & 20 & 0.25 & 0.1267 & \text{0.1032(1)} & \text{0.54(3)} & \text{0.57(3)} & 0.97 & \text{0.8k} \\
 40 & 20 & 0.25 & 0.125 & \text{0.1700(2)} & \text{0.55(2)} & \text{0.55(2)} & 0.92 & \text{0.8k} \\
 40 & 20 & 0.25 & 0.12 & \text{0.3823(2)} & \text{0.801(2)} & \text{0.801(2)} & 0.91 & \text{2k} \\
 40 & 20 & 0.25 & 0.115 & \text{0.7534(6)} & \text{1.373(7)} & \text{1.374(7)} & 0.92 & \text{2k} \\
\end{tabular}
\caption{Same as Table \ref{tbl:simparamnt24} for $N_t=40$, $N_s=20$.}
\label{tbl:simparamnt40ns20}
\end{table}

\begin{table}[h]
\centering
\begin{tabular}{c | c | c | c | c | c | c | c | r}
$N_{t}$ & $N_{s}$ & $\beta_{L}$ & $\kappa$ & $m_q$ & $m_{\pi}$ & $m_{\rho}$ & acc. & stat. \\\hline
 40 & 30 & -0.25 & 0.1309 & \text{0.0209(1)} & \text{0.32(2)} & \text{0.28(2)} & 0.91 & \text{1.9k} \\
 40 & 30 & -0.25 & 0.1302 & \text{0.0489(1)} & \text{0.40(2)} & \text{0.39(2)} & 0.92 & \text{0.8k} \\
 40 & 30 & -0.25 & 0.129 & \text{0.1015(2)} & \text{0.39(3)} & \text{0.38(3)} & 0.75 & \text{0.3k} \\
 40 & 30 & -0.25 & 0.1277 & \text{0.1615(2)} & \text{0.401(3)} & \text{0.396(3)} & 0.74 & \text{0.2k} \\
 40 & 30 & -0.25 & 0.1263 & \text{0.2270(2)} & \text{0.535(8)} & \text{0.535(8)} & 0.77 & \text{1k} \\
 40 & 30 & -0.25 & 0.125 & \text{0.3018(2)} & \text{0.670(1)} & \text{0.670(1)} & 0.79 & \text{2k} \\
 40 & 30 & -0.25 & 0.123 & \text{0.4557(6)} & \text{0.981(6)} & \text{0.981(6)} & 0.76 & \text{2k} \\
 40 & 30 & 0.001 & 0.1299 & \text{0.0176(1)} & \text{0.37(2)} & \text{0.38(2)} & 0.94 & \text{1.7k} \\
 40 & 30 & 0.001 & 0.129 & \text{0.0523(1)} & \text{0.32(2)} & \text{0.29(2)} & 0.79 & \text{1k} \\
 40 & 30 & 0.001 & 0.1278 & \text{0.0991(1)} & \text{0.40(2)} & \text{0.38(2)} & 0.94 & \text{1k} \\
 40 & 30 & 0.001 & 0.125 & \text{0.2185(2)} & \text{0.514(3)} & \text{0.509(3)} & 0.92 & \text{0.5k} \\
 40 & 30 & 0.001 & 0.12 & \text{0.5091(4)} & \text{1.040(5)} & \text{1.040(5)} & 0.8 & \text{2k} \\
 40 & 30 & 0.25 & 0.129 & \text{0.0155(1)} & \text{0.29(2)} & \text{0.25(2)} & 0.81 & \text{1k} \\
 40 & 30 & 0.25 & 0.1281 & \text{0.0503(1)} & \text{0.36(2)} & \text{0.34(3)} & 0.94 & \text{0.9k} \\
 40 & 30 & 0.25 & 0.1267 & \text{0.0994(1)} & \text{0.29(1)} & \text{0.27(1)} & 0.94 & \text{0.5k} \\
 40 & 30 & 0.25 & 0.125 & \text{0.1659(2)} & \text{0.393(2)} & \text{0.387(2)} & 0.81 & \text{0.4k} \\
 40 & 30 & 0.25 & 0.115 & \text{0.7552(8)} & \text{1.398(6)} & \text{1.398(6)} & 0.83 & \text{2k} \\
  \end{tabular}
\caption{Same as Table \ref{tbl:simparamnt24} for $N_t=40$, $N_s=30$.}
\label{tbl:simparamnt40ns30}
\end{table}

\begin{table}[h]
\centering
\begin{tabular}{c | c | c | c | c | c | c | c | r}
$N_{t}$ & $N_{s}$ & $\beta_{L}$ & $\kappa$ & $m_q$ & $m_{\pi}$ & $m_{\rho}$ & acc. & stat. \\\hline
40 & 40 & -0.25 & 0.1309 & \text{0.0203(1)} & \text{0.25(3)} & \text{0.22(3)} & 0.91 & \text{0.9k} \\
 40 & 40 & -0.25 & 0.129 & \text{0.1005(1)} & \text{0.262(1)} & \text{0.255(1)} & 0.88 & \text{0.8k} \\
 40 & 40 & -0.25 & 0.1277 & \text{0.1610(1)} & \text{0.375(1)} & \text{0.374(1)} & 0.88 & \text{2k} \\
 40 & 40 & -0.25 & 0.1263 & \text{0.2272(1)} & \text{0.523(2)} & \text{0.521(2)} & 0.9 & \text{2k} \\
 40 & 40 & -0.25 & 0.125 & \text{0.3023(2)} & \text{0.672(1)} & \text{0.672(1)} & 0.89 & \text{2k} \\
 40 & 40 & -0.25 & 0.123 & \text{0.4561(5)} & \text{0.974(2)} & \text{0.975(2)} & 0.88 & \text{2k} \\
 40 & 40 & 0.001 & 0.1299 & \text{0.0173(1)} & \text{0.25(2)} & \text{0.21(3)} & 0.9 & \text{0.8k} \\
 40 & 40 & 0.001 & 0.129 & \text{0.0521(1)} & \text{0.25(2)} & \text{0.23(2)} & 0.89 & \text{1.1k} \\
 40 & 40 & 0.001 & 0.125 & \text{0.2184(1)} & \text{0.492(1)} & \text{0.491(1)} & 0.9 & \text{2k} \\
 40 & 40 & 0.001 & 0.12 & \text{0.5100(4)} & \text{1.046(5)} & \text{1.045(5)} & 0.9 & \text{2k} \\
 40 & 40 & 0.25 & 0.129 & \text{0.0151(1)} & \text{0.20(2)} & \text{0.17(2)} & 0.91 & \text{0.9k} \\
 40 & 40 & 0.25 & 0.125 & \text{0.1662(1)} & \text{0.377(2)} & \text{0.376(2)} & 0.91 & \text{2k} \\
 40 & 40 & 0.25 & 0.12 & \text{0.3868(2)} & \text{0.807(2)} & \text{0.807(2)} & 0.92 & \text{2k} \\
 40 & 40 & 0.25 & 0.115 & \text{0.758(2)} & \text{1.41(1)} & \text{1.42(1)} & 0.92 & \text{2k} \\
\end{tabular}
\caption{Same as  Table \ref{tbl:simparamnt24} for $N_t=40$, $N_s=40$.}
\label{tbl:simparamnt40ns40}
\end{table}

\begin{table}[h]
\centering
\begin{tabular}{c | c | c | c | c | c | c | c | r}
$N_{t}$ & $N_{s}$ & $\beta_{L}$ & $\kappa$ & $m_q$ & $m_{\pi}$ & $m_{\rho}$ & acc. & stat. \\\hline
 48 & 24 & -0.25 & 0.1309 & \text{0.0214(1)} & \text{0.44(2)} & \text{0.45(2)} & 0.94 & \text{2k} \\
 48 & 24 & -0.25 & 0.1302 & \text{0.0497(1)} & \text{0.43(2)} & \text{0.40(2)} & 0.95 & \text{1.6k} \\
 48 & 24 & -0.25 & 0.129 & \text{0.1027(2)} & \text{0.41(2)} & \text{0.43(3)} & 0.82 & \text{2k} \\
 48 & 24 & -0.25 & 0.1277 & \text{0.1608(1)} & \text{0.435(7)} & \text{0.443(7)} & 0.81 & \text{2k} \\
 48 & 24 & -0.25 & 0.1263 & \text{0.2245(1)} & \text{0.515(1)} & \text{0.516(2)} & 0.81 & \text{2k} \\
 48 & 24 & -0.25 & 0.125 & \text{0.2991(3)} & \text{0.674(3)} & \text{0.674(3)} & 0.83 & \text{2k} \\
 48 & 24 & -0.25 & 0.123 & \text{0.4539(4)} & \text{0.990(8)} & \text{0.990(9)} & 0.82 & \text{2k} \\
 48 & 24 & 0.001 & 0.1299 & \text{0.0182(1)} & \text{0.41(2)} & \text{0.42(2)} & 0.95 & \text{2k} \\
 48 & 24 & 0.001 & 0.129 & \text{0.0535(2)} & \text{0.45(2)} & \text{0.41(2)} & 0.84 & \text{2k} \\
 48 & 24 & 0.001 & 0.1278 & \text{0.1003(1)} & \text{0.49(1)} & \text{0.47(2)} & 0.95 & \text{1.8k} \\
 48 & 24 & 0.001 & 0.125 & \text{0.2168(2)} & \text{0.508(2)} & \text{0.505(2)} & 0.84 & \text{2k} \\
 48 & 24 & 0.001 & 0.12 & \text{0.5063(4)} & \text{1.037(5)} & \text{1.037(5)} & 0.85 & \text{2k} \\
 48 & 24 & 0.25 & 0.129 & \text{0.0161(1)} & \text{0.42(2)} & \text{0.43(3)} & 0.86 & \text{2k} \\
 48 & 24 & 0.25 & 0.1281 & \text{0.0510(1)} & \text{0.42(2)} & \text{0.39(2)} & 0.95 & \text{1.8k} \\
 48 & 24 & 0.25 & 0.1267 & \text{0.1008(2)} & \text{0.385(7)} & \text{0.362(7)} & 0.95 & \text{1.8k} \\
 48 & 24 & 0.25 & 0.125 & \text{0.1648(1)} & \text{0.382(2)} & \text{0.384(2)} & 0.86 & \text{2k} \\
 48 & 24 & 0.25 & 0.12 & \text{0.3821(2)} & \text{0.798(1)} & \text{0.798(1)} & 0.87 & \text{2k} \\
 48 & 24 & 0.25 & 0.115 & \text{0.7508(8)} & \text{1.384(8)} & \text{1.385(8)} & 0.89 & \text{2k} \\
 \end{tabular}
\caption{Same as  Table \ref{tbl:simparamnt24} for $N_t=48$, $N_s=24$.}
\label{tbl:simparamnt48ns24}
\end{table}

\begin{table}[h]
\centering
\begin{tabular}{c | c | c | c | c | c | c | c | r}
$N_{t}$ & $N_{s}$ & $\beta_{L}$ & $\kappa$ & $m_q$ & $m_{\pi}$ & $m_{\rho}$ & acc. & stat. \\\hline
 48 & 36 & -0.25 & 0.1309 & \text{0.0205(1)} & \text{0.28(2)} & \text{0.25(3)} & 0.88 & \text{1k} \\
 48 & 36 & -0.25 & 0.129 & \text{0.0999(1)} & \text{0.238(1)} & \text{0.237(1)} & 0.89 & \text{2k} \\
 48 & 36 & -0.25 & 0.1277 & \text{0.1602(1)} & \text{0.372(1)} & \text{0.372(1)} & 0.9 & \text{2k} \\
 48 & 36 & -0.25 & 0.1263 & \text{0.2260(1)} & \text{0.513(1)} & \text{0.513(1)} & 0.88 & \text{2k} \\
 48 & 36 & -0.25 & 0.125 & \text{0.3006(2)} & \text{0.670(2)} & \text{0.670(2)} & 0.89 & \text{2k} \\
 48 & 36 & -0.25 & 0.123 & \text{0.4538(5)} & \text{0.973(3)} & \text{0.973(3)} & 0.89 & \text{2k} \\
 48 & 36 & 0.001 & 0.1299 & \text{0.0173(1)} & \text{0.30(3)} & \text{0.28(3)} & 0.9 & \text{2.4k} \\
 48 & 36 & 0.001 & 0.129 & \text{0.0520(1)} & \text{0.24(2)} & \text{0.22(2)} & 0.91 & \text{0.6k} \\
 48 & 36 & 0.001 & 0.125 & \text{0.2175(1)} & \text{0.488(1)} & \text{0.487(1)} & 0.91 & \text{2k} \\
 48 & 36 & 0.001 & 0.12 & \text{0.5073(3)} & \text{1.038(3)} & \text{1.039(3)} & 0.91 & \text{2k} \\
 48 & 36 & 0.25 & 0.129 & \text{0.0152(1)} & \text{0.25(2)} & \text{0.21(2)} & 0.92 & \text{1.6k} \\
 48 & 36 & 0.25 & 0.125 & \text{0.1653(1)} & \text{0.376(1)} & \text{0.375(1)} & 0.91 & \text{2k} \\
 48 & 36 & 0.25 & 0.12 & \text{0.3851(2)} & \text{0.803(1)} & \text{0.803(1)} & 0.92 & \text{2k} \\
 48 & 36 & 0.25 & 0.115 & \text{0.754(1)} & \text{1.42(2)} & \text{1.42(2)} & 0.92 & \text{2k} \\
 \end{tabular}
\caption{Same as  Table \ref{tbl:simparamnt24} for $N_t=48$, $N_s=36$.}
\label{tbl:simparamnt48ns36}
\end{table}

\begin{table}[h]
\centering
\begin{tabular}{c | c | c | c | c | c | c | c | r}
$N_{t}$ & $N_{s}$ & $\beta_{L}$ & $\kappa$ & $m_q$ & $m_{\pi}$ & $m_{\rho}$ & acc. & stat. \\\hline
 48 & 48 & -0.25 & 0.1309 & \text{0.0202(1)} & \text{0.16(2)} & \text{0.13(3)} & 0.92 & \text{1.4k} \\
 48 & 48 & -0.25 & 0.129 & \text{0.1001(1)} & \text{0.239(1)} & \text{0.238(1)} & 0.91 & \text{2.9k} \\
 48 & 48 & -0.25 & 0.1277 & \text{0.1608(1)} & \text{0.373(1)} & \text{0.372(1)} & 0.9 & \text{3k} \\
 48 & 48 & -0.25 & 0.1263 & \text{0.2266(1)} & \text{0.516(1)} & \text{0.516(1)} & 0.91 & \text{3.2k} \\
 48 & 48 & -0.25 & 0.125 & \text{0.3013(2)} & \text{0.672(1)} & \text{0.672(1)} & 0.91 & \text{3.3k} \\
 48 & 48 & -0.25 & 0.123 & \text{0.4546(3)} & \text{0.976(2)} & \text{0.976(2)} & 0.91 & \text{3.5k} \\
 48 & 48 & 0.001 & 0.1299 & \text{0.0170(1)} & \text{0.19(2)} & \text{0.16(3)} & 0.92 & \text{2.1k} \\
 48 & 48 & 0.001 & 0.129 & \text{0.0517(1)} & \text{0.173(4)} & \text{0.158(4)} & 0.93 & \text{1.3k} \\
 48 & 48 & 0.001 & 0.125 & \text{0.2179(1)} & \text{0.492(1)} & \text{0.491(1)} & 0.92 & \text{3.2k} \\
 48 & 48 & 0.001 & 0.12 & \text{0.5074(3)} & \text{1.052(3)} & \text{1.053(3)} & 0.93 & \text{3.6k} \\
 48 & 48 & 0.25 & 0.129 & \text{0.0151(1)} & \text{0.18(3)} & \text{0.15(3)} & 0.91 & \text{1.9k} \\
 48 & 48 & 0.25 & 0.125 & \text{0.1658(1)} & \text{0.373(1)} & \text{0.373(1)} & 0.93 & \text{2.9k} \\
 48 & 48 & 0.25 & 0.12 & \text{0.3853(1)} & \text{0.810(3)} & \text{0.810(3)} & 0.93 & \text{3.6k} \\
 48 & 48 & 0.25 & 0.115 & \text{0.7534(7)} & \text{1.42(2)} & \text{1.42(2)} & 0.94 & \text{3.4k} \\
\end{tabular}
\caption{Same as  Table \ref{tbl:simparamnt24} for $N_t=48$, $N_s=48$.}
\label{tbl:simparamnt48ns48}
\end{table}


\newpage

\begin{thebibliography}{99}
\bibitem{Sannino:2004qp}
F.~Sannino and K.~Tuominen,
Phys. Rev. D \textbf{71} (2005), 051901
doi:\href{https://doi.org/10.1103/PhysRevD.71.051901}{10.1103/PhysRevD.71.051901}
[arXiv:\href{https://arxiv.org/abs/hep-ph/0405209}{hep-ph/0405209} [hep-ph]].

\bibitem{Hill:2002ap}
C.~T.~Hill and E.~H.~Simmons,
Phys. Rept. \textbf{381} (2003), 235-402
[erratum: Phys. Rept. \textbf{390} (2004), 553-554]
doi:\href{https://doi.org/10.1016/S0370-1573(03)00140-6}{10.1016/S0370-1573(03)00140-6}
[arXiv:\href{https://arxiv.org/abs/hep-ph/0203079}{hep-ph/0203079} [hep-ph]].

\bibitem{Dietrich:2005jn}
D.~D.~Dietrich, F.~Sannino and K.~Tuominen,
Phys. Rev. D \textbf{72} (2005), 055001
doi:\href{https://doi.org/10.1103/PhysRevD.72.055001}{10.1103/PhysRevD.72.055001}
[arXiv:\href{https://arxiv.org/abs/hep-ph/0505059}{hep-ph/0505059} [hep-ph]].

\bibitem{Arbey:2015exa}
A.~Arbey, G.~Cacciapaglia, H.~Cai, A.~Deandrea, S.~Le Corre and F.~Sannino,
Phys. Rev. D \textbf{95} (2017) no.1, 015028
doi:\href{https://doi.org/10.1103/PhysRevD.95.015028}{10.1103/PhysRevD.95.015028}
[arXiv:\href{https://arxiv.org/abs/1502.04718}{1502.04718} [hep-ph]].

\bibitem{Karavirta:2011zg}
T.~Karavirta, J.~Rantaharju, K.~Rummukainen and K.~Tuominen,
JHEP \textbf{05} (2012), 003
doi:\href{https://doi.org/10.1007/JHEP05(2012)003}{10.1007/JHEP05(2012)003}
[arXiv:\href{https://arxiv.org/abs/1111.4104}{1111.4104} [hep-lat]].

\bibitem{Leino:2017lpc}
V.~Leino, J.~Rantaharju, T.~Rantalaiho, K.~Rummukainen, J.~M.~Suorsa and K.~Tuominen,
Phys. Rev. D \textbf{95} (2017) no.11, 114516
doi:\href{https://doi.org/10.1103/PhysRevD.95.114516}{10.1103/PhysRevD.95.114516}
[arXiv:\href{https://arxiv.org/abs/1701.04666}{1701.04666} [hep-lat]].

\bibitem{Leino:2017hgm}
V.~Leino, K.~Rummukainen, J.~M.~Suorsa, K.~Tuominen and S.~T\"ahtinen,
Phys. Rev. D \textbf{97} (2018) no.11, 114501
doi:\href{https://doi.org/10.1103/PhysRevD.97.114501}{10.1103/PhysRevD.97.114501}
[arXiv:\href{https://arxiv.org/abs/1707.04722}{1707.04722} [hep-lat]].

\bibitem{Leino:2018qvq}
V.~Leino, K.~Rummukainen and K.~Tuominen,
Phys. Rev. D \textbf{98} (2018) no.5, 054503
doi:\href{https://doi.org/10.1103/PhysRevD.98.054503}{10.1103/PhysRevD.98.054503}
[arXiv:\href{https://arxiv.org/abs/1804.02319}{1804.02319} [hep-lat]].

\bibitem{Amato:2018nvj}
A.~Amato, V.~Leino, K.~Rummukainen, K.~Tuominen and S.~T\"ahtinen,
[arXiv:\href{https://arxiv.org/abs/1806.07154}{1806.07154} [hep-lat]].

\bibitem{Hietanen:2008mr}
A.~J.~Hietanen, J.~Rantaharju, K.~Rummukainen and K.~Tuominen,
JHEP \textbf{05} (2009), 025
doi:\href{https://doi.org/10.1088/1126-6708/2009/05/025}{10.1088/1126-6708/2009/05/025}
[arXiv:\href{https://arxiv.org/abs/0812.1467}{0812.1467} [hep-lat]].

\bibitem{Hietanen:2009az}
A.~J.~Hietanen, K.~Rummukainen and K.~Tuominen,
Phys. Rev. D \textbf{80} (2009), 094504
doi:\href{https://doi.org/10.1103/PhysRevD.80.094504}{10.1103/PhysRevD.80.094504}
[arXiv:\href{https://arxiv.org/abs/0904.0864}{0904.0864} [hep-lat]].

\bibitem{DelDebbio:2008zf}
L.~Del Debbio, A.~Patella and C.~Pica,
Phys. Rev. D \textbf{81} (2010), 094503
doi:\href{https://doi.org/10.1103/PhysRevD.81.094503}{10.1103/PhysRevD.81.094503}
[arXiv:\href{https://arxiv.org/abs/0805.2058}{0805.2058} [hep-lat]].

\bibitem{DelDebbio:2009fd}
L.~Del Debbio, B.~Lucini, A.~Patella, C.~Pica and A.~Rago,
Phys. Rev. D \textbf{80} (2009), 074507
doi:\href{https://doi.org/10.1103/PhysRevD.80.074507}{10.1103/PhysRevD.80.074507}
[arXiv:\href{https://arxiv.org/abs/0907.3896}{0907.3896} [hep-lat]].

\bibitem{DelDebbio:2010hu}
L.~Del Debbio, B.~Lucini, A.~Patella, C.~Pica and A.~Rago,
Phys. Rev. D \textbf{82} (2010), 014509
doi:\href{https://doi.org/10.1103/PhysRevD.82.014509}{10.1103/PhysRevD.82.014509}
[arXiv:\href{https://arxiv.org/abs/1004.3197}{1004.3197} [hep-lat]].

\bibitem{Bursa:2011ru}
F.~Bursa, L.~Del Debbio, D.~Henty, E.~Kerrane, B.~Lucini, A.~Patella, C.~Pica, T.~Pickup and A.~Rago,
Phys. Rev. D \textbf{84} (2011), 034506
doi:\href{https://doi.org/10.1103/PhysRevD.84.034506}{10.1103/PhysRevD.84.034506}
[arXiv:\href{https://arxiv.org/abs/1104.4301}{1104.4301} [hep-lat]].

\bibitem{DeGrand:2011qd}
T.~DeGrand, Y.~Shamir and B.~Svetitsky,
Phys. Rev. D \textbf{83} (2011), 074507
doi:\href{https://doi.org/10.1103/PhysRevD.83.074507}{10.1103/PhysRevD.83.074507}
[arXiv:\href{https://arxiv.org/abs/1102.2843}{1102.2843} [hep-lat]].

\bibitem{Rantaharju:2015yva}
J.~Rantaharju, T.~Rantalaiho, K.~Rummukainen and K.~Tuominen,
Phys. Rev. D \textbf{93} (2016) no.9, 094509
doi:\href{https://doi.org/10.1103/PhysRevD.93.094509}{10.1103/PhysRevD.93.094509}
[arXiv:\href{https://arxiv.org/abs/1510.03335}{1510.03335} [hep-lat]].

\bibitem{DelDebbio:2015byq}
L.~Del Debbio, B.~Lucini, A.~Patella, C.~Pica and A.~Rago,
Phys. Rev. D \textbf{93} (2016) no.5, 054505
doi:\href{https://doi.org/10.1103/PhysRevD.93.054505}{10.1103/PhysRevD.93.054505}
[arXiv:\href{https://arxiv.org/abs/1512.08242}{1512.08242} [hep-lat]].

\bibitem{Fodor:2011tu}
Z.~Fodor, K.~Holland, J.~Kuti, D.~Nogradi, C.~Schroeder, K.~Holland, J.~Kuti, D.~Nogradi and C.~Schroeder,
Phys. Lett. B \textbf{703} (2011), 348-358
doi:\href{https://doi.org/10.1016/j.physletb.2011.07.037}{10.1016/j.physletb.2011.07.037}
[arXiv:\href{https://arxiv.org/abs/1104.3124}{1104.3124} [hep-lat]].

\bibitem{Hasenfratz:2014rna}
A.~Hasenfratz, D.~Schaich and A.~Veernala,
JHEP \textbf{06} (2015), 143
doi:\href{https://doi.org/10.1007/JHEP06(2015)143}{10.1007/JHEP06(2015)143}
[arXiv:\href{https://arxiv.org/abs/1410.5886}{1410.5886} [hep-lat]].

\bibitem{Fodor:2015baa}
Z.~Fodor, K.~Holland, J.~Kuti, S.~Mondal, D.~Nogradi and C.~H.~Wong,
JHEP \textbf{06} (2015), 019
doi:\href{https://doi.org/10.1007/JHEP06(2015)019}{10.1007/JHEP06(2015)019}
[arXiv:\href{https://arxiv.org/abs/1503.01132}{1503.01132} [hep-lat]].

\bibitem{Fodor:2017gtj}
Z.~Fodor, K.~Holland, J.~Kuti, D.~Nogradi and C.~H.~Wong,
Phys. Lett. B \textbf{779} (2018), 230-236
doi:\href{https://doi.org/10.1016/j.physletb.2018.02.008}{10.1016/j.physletb.2018.02.008}
[arXiv:\href{https://arxiv.org/abs/1710.09262}{1710.09262} [hep-lat]].

\bibitem{LatticeStrongDynamics:2018hun}
T.~Appelquist \textit{et al.} [Lattice Strong Dynamics],
Phys. Rev. D \textbf{99} (2019) no.1, 014509
doi:\href{https://doi.org/10.1103/PhysRevD.99.014509}{10.1103/PhysRevD.99.014509}
[arXiv:\href{https://arxiv.org/abs/1807.08411}{1807.08411} [hep-lat]].

\bibitem{Hasenfratz:2019dpr}
A.~Hasenfratz, C.~Rebbi and O.~Witzel,
Phys. Rev. D \textbf{100} (2019) no.11, 114508
doi:\href{https://doi.org/10.1103/PhysRevD.100.114508}{10.1103/PhysRevD.100.114508}
[arXiv:\href{https://arxiv.org/abs/1909.05842}{1909.05842} [hep-lat]].

\bibitem{Hasenfratz:2020ess}
A.~Hasenfratz, C.~Rebbi and O.~Witzel,
Phys. Rev. D \textbf{101} (2020) no.11, 114508
doi:\href{https://doi.org/10.1103/PhysRevD.101.114508}{10.1103/PhysRevD.101.114508}
[arXiv:\href{https://arxiv.org/abs/2004.00754}{2004.00754} [hep-lat]].

\bibitem{Shamir:2008pb}
Y.~Shamir, B.~Svetitsky and T.~DeGrand,
Phys. Rev. D \textbf{78} (2008), 031502
doi:\href{https://doi.org/10.1103/PhysRevD.78.031502}{10.1103/PhysRevD.78.031502}
[arXiv:\href{https://arxiv.org/abs/0803.1707}{0803.1707} [hep-lat]].

\bibitem{DeGrand:2008kx}
T.~DeGrand, Y.~Shamir and B.~Svetitsky,
Phys. Rev. D \textbf{79} (2009), 034501
doi:\href{https://doi.org/10.1103/PhysRevD.79.034501}{10.1103/PhysRevD.79.034501}
[arXiv:\href{https://arxiv.org/abs/0812.1427}{0812.1427} [hep-lat]].

\bibitem{Fodor:2009ar}
Z.~Fodor, K.~Holland, J.~Kuti, D.~Nogradi and C.~Schroeder,
JHEP \textbf{11} (2009), 103
doi:\href{https://doi.org/10.1088/1126-6708/2009/11/103}{10.1088/1126-6708/2009/11/103}
[arXiv:\href{https://arxiv.org/abs/0908.2466}{0908.2466} [hep-lat]].

\bibitem{DeGrand:2010na}
T.~DeGrand, Y.~Shamir and B.~Svetitsky,
Phys. Rev. D \textbf{82} (2010), 054503
doi:\href{https://doi.org/10.1103/PhysRevD.82.054503}{10.1103/PhysRevD.82.054503}
[arXiv:\href{https://arxiv.org/abs/1006.0707}{1006.0707} [hep-lat]].

\bibitem{Fodor:2015zna}
Z.~Fodor, K.~Holland, J.~Kuti, S.~Mondal, D.~Nogradi and C.~H.~Wong,
JHEP \textbf{09} (2015), 039
doi:\href{https://doi.org/10.1007/JHEP09(2015)039}{10.1007/JHEP09(2015)039}
[arXiv:\href{https://arxiv.org/abs/1506.06599}{1506.06599} [hep-lat]].

\bibitem{DeGrand:2012qa}
T.~DeGrand, Y.~Shamir and B.~Svetitsky,
Phys. Rev. D \textbf{85} (2012), 074506
doi:\href{https://doi.org/10.1103/PhysRevD.85.074506}{10.1103/PhysRevD.85.074506}
[arXiv:\href{https://arxiv.org/abs/1202.2675}{1202.2675} [hep-lat]].

\bibitem{DeGrand:2015lna}
T.~DeGrand, Y.~Liu, E.~T.~Neil, Y.~Shamir and B.~Svetitsky,
Phys. Rev. D \textbf{91} (2015), 114502
doi:\href{https://doi.org/10.1103/PhysRevD.91.114502}{10.1103/PhysRevD.91.114502}
[arXiv:\href{https://arxiv.org/abs/1501.05665}{1501.05665} [hep-lat]].

\bibitem{Leino:2019qwk}
V.~Leino, T.~Rindlisbacher, K.~Rummukainen, F.~Sannino and K.~Tuominen,
Phys. Rev. D \textbf{101} (2020) no.7, 074508
doi:\href{https://doi.org/10.1103/PhysRevD.101.074508}{10.1103/PhysRevD.101.074508}
[arXiv:\href{https://arxiv.org/abs/1908.04605}{1908.04605} [hep-lat]].

\bibitem{Jegerlehner:1998zg}
F.~Jegerlehner and O.~V.~Tarasov,
Nucl. Phys. B \textbf{549} (1999), 481-498
doi:\href{https://doi.org/10.1016/S0550-3213(99)00141-8}{10.1016/S0550-3213(99)00141-8}
[arXiv:\href{https://arxiv.org/abs/hep-ph/9809485}{hep-ph/9809485} [hep-ph]].

\bibitem{Capitani:2006ni}
S.~Capitani, S.~Durr and C.~Hoelbling,
JHEP \textbf{11} (2006), 028
doi:\href{https://doi.org/10.1088/1126-6708/2006/11/028}{10.1088/1126-6708/2006/11/028}
[arXiv:\href{https://arxiv.org/abs/hep-lat/0607006}{hep-lat/0607006} [hep-lat]].

\bibitem{Brower:1995vx}
R.~C.~Brower, T.~Ivanenko, A.~R.~Levi and K.~N.~Orginos,
Nucl. Phys. B \textbf{484} (1997), 353-374
doi:\href{https://doi.org/10.1016/S0550-3213(96)00579-2}{10.1016/S0550-3213(96)00579-2}
[arXiv:\href{https://arxiv.org/abs/hep-lat/9509012}{hep-lat/9509012} [hep-lat]].

\bibitem{Hasenfratz:1993az}
A.~Hasenfratz and T.~A.~DeGrand,
Phys. Rev. D \textbf{49} (1994), 466-473
doi:\href{https://doi.org/10.1103/PhysRevD.49.466}{10.1103/PhysRevD.49.466}
[arXiv:\href{https://arxiv.org/abs/hep-lat/9304001}{hep-lat/9304001} [hep-lat]].

\bibitem{Blum:1994xb}
T.~Blum, C.~E.~DeTar, U.~M.~Heller, L.~Karkkainen, K.~Rummukainen and D.~Toussaint,
Nucl. Phys. B \textbf{442} (1995), 301-316
doi:\href{https://doi.org/10.1016/0550-3213(95)00137-9}{10.1016/0550-3213(95)00137-9}
[arXiv:\href{https://arxiv.org/abs/hep-lat/9412038}{hep-lat/9412038} [hep-lat]].

\bibitem{deForcrand:2012vh}
P.~de Forcrand, S.~Kim and W.~Unger,
JHEP \textbf{02} (2013), 051
doi:\href{https://doi.org/10.1007/JHEP02(2013)051}{10.1007/JHEP02(2013)051}
[arXiv:\href{https://arxiv.org/abs/1208.2148}{1208.2148} [hep-lat]].

\bibitem{Bolder:2000un}
B.~Bolder, T.~Struckmann, G.~S.~Bali, N.~Eicker, T.~Lippert, B.~Orth, K.~Schilling and P.~Ueberholz,
Phys. Rev. D \textbf{63} (2001), 074504
doi:\href{https://doi.org/10.1103/PhysRevD.63.074504}{10.1103/PhysRevD.63.074504}
[arXiv:\href{https://arxiv.org/abs/hep-lat/0005018}{hep-lat/0005018} [hep-lat]].

\end{thebibliography}
\end{document}